\documentclass[%
 reprint,onecolumn,
superscriptaddress,
nofootinbib,
 amsmath,amssymb,
 aps,
]{revtex4}

\usepackage{graphicx}% Include figure files
\usepackage{dcolumn}% Align table columns on decimal point
\usepackage{bm}% bold math
\usepackage{xcolor}
\usepackage{hyperref}
\usepackage{cleveref}

\begin{document}

\newtheorem{theorem}{Theorem}
\newtheorem{definition}{Definition}
\newtheorem{lemma}{Lemma}
\newtheorem{proposition}{Proposition}
\newtheorem{remark}{Remark}
\newtheorem{corollary}{Corollary}
\newtheorem{example}{Example}

%\preprint{APS/123-QED}

\title{Stochastic dynamics with multiplicative dichotomic noise: heterogeneous telegrapher's equation, anomalous crossovers and resetting}% Force line breaks with \\

% please add all authors name and surname separated by comma
\author{Trifce Sandev}
\email{trifce.sandev@manu.edu.mk}
\affiliation{Research Center for Computer Science and Information Technologies, Macedonian Academy of Sciences and Arts, Bul. Krste Misirkov 2, 1000 Skopje, Macedonia}
\affiliation{Institute of Physics \& Astronomy, University of Potsdam, D-14776 Potsdam-Golm, Germany}
\affiliation{Institute of Physics, Faculty of Natural Sciences and Mathematics,
Ss.~Cyril and Methodius University, Arhimedova 3, 1000 Skopje, Macedonia}

\author{Ljupco Kocarev}
\email{lkocarev@manu.edu.mk}
\affiliation{Research Center for Computer Science and Information Technologies, Macedonian Academy of Sciences and Arts, Bul. Krste Misirkov 2, 1000 Skopje, Macedonia}
\affiliation{Faculty of Computer Science and Engineering, Ss. Cyril and Methodius University, \\ PO Box 393, 1000 Skopje, Macedonia}

\author{Ralf Metzler}
\email{rmetzler@uni-potsdam.de}
\affiliation{Institute of Physics \& Astronomy, University of Potsdam, D-14776 Potsdam-Golm, Germany}
\affiliation{Asia Pacific Center for Theoretical Physics, Pohang 37673, Republic of Korea}

\author{Aleksei Chechkin}
\email{chechkin@uni-potsdam.de}
\affiliation{Institute of Physics \& Astronomy, University of Potsdam, D-14776 Potsdam-Golm, Germany}
\affiliation{Faculty of Pure and Applied Mathematics, Hugo Steinhaus Center, Wroclaw University of Science and Technology, Wyspianskiego 27, 50-370 Wroclaw, Poland}
\affiliation{Akhiezer Institute for Theoretical Physics National Science Center ``Kharkiv Institute of Physics and Technology'', Akademichna 1, Kharkiv 61108, Ukraine}

\date{\today}% It is always \today, today,

\begin{abstract}
We analyze diffusion processes with finite propagation speed in a non-homogeneous medium in terms of the heterogeneous telegrapher's equation. In the diffusion limit of infinite-velocity propagation we recover the results for the heterogeneous diffusion process. The heterogeneous telegrapher's process exhibits a rich variety of diffusion regimes including hyperdiffusion, ballistic motion, superdiffusion, normal diffusion and subdiffusion, and different crossover dynamics characteristic for complex systems in which anomalous diffusion is observed. The anomalous diffusion exponent in the short time limit is twice the exponent in the long time limit, in accordance to the crossover dynamics from ballistic diffusion to normal diffusion in the standard telegrapher's process. We also analyze the finite-velocity heterogeneous diffusion process in presence of stochastic Poissonian resetting. We show that the system reaches a non-equilibrium stationary state. The transition to this non-equilibrium steady state is analysed in terms of the large deviation function.
\end{abstract}
\maketitle

\section{Introduction}

It is well known that the Green's function of the classical diffusion equation, the Gaussian distribution, has non-zero values for any $x$ at $t>0$, which means that some of the particles move with an arbitrarily chosen large velocity. To avoid this unphysical property, a finite-velocity diffusion process governed by the so-called telegrapher's (or Cattaneo) equation was introduced, and a corresponding persistent random walk model was proposed. Historically, the telegrapher's equation has been derived by Heaviside for a voltage $u$ along a lossy transmission line in electrodynamic theory~\cite{heaviside},
\begin{align}\label{classical cattaneo eq}
\tau\frac{\partial^{2}}{\partial t^{2}}u(x,t)+\frac{\partial}{\partial t}u(x,t)=D\frac{\partial^{2}}{\partial x^{2}}u(x,t).
\end{align}
Here $\tau$ is a time parameter, and $D$ is the diffusion coefficient, which relates to a finite propagation velocity $v=\sqrt{D/\tau}$. Contrary to the diffusion equation which is parabolic, the telegrapher's equation is a hyperbolic partial differential equation. A simple interpretation of this process is that the probability flux is delayed over time by the interval $\tau$ with respect to the probability gradient, $J(x,t+\tau)=-D\frac{\partial}{\partial x}u(x,t)$. Assuming $\tau\ll t$, then 
\begin{align}\label{gradient}
    J(x,t)+\tau\frac{\partial}{\partial t}J(x,t)=-D\frac{\partial}{\partial x}u(x,t).
\end{align} 
This equation was proposed by Cattaneo in 1948 \cite{cattaneo1} (see also~\cite{cattaneo2,new}) to extend the standard constitutive relation. Combining this equation with continuity equation 
\begin{align}\label{continuity}
    \frac{\partial}{\partial t}u(x,t)=-\frac{\partial}{\partial x}J(x,t),
\end{align}
one obtains the telegrapher's equation~(\ref{classical cattaneo eq}) that is often alternatively referred to as Cattaneo equation. The persistent random walk was suggested first by F\"{u}rth \cite{furth} and Taylor \cite{taylor}, who considered it as a suitable model for transport in turbulent diffusion, while Goldstein gave solutions of various forms of the telegrapher's equation~\cite{goldstein} (see also \cite{Jagher}). The telegrapher's equation can be considered as a particular case of a spatio-temporally coupled L\'evy walk model with exponential waiting time probability density~\cite{zumofen,klafter,levywalkreview}. Extended Poisson-Kac theory provides a unifying framework for stochastic processes with finite propagation velocity and was developed recently~\cite{klages}. The telegrapher's equation was also used to study finite-velocity diffusion on a comb~\cite{epl} and in random media~\cite{jsp}, as well as the telegraph processes with random velocities~\cite{jstor}. Fractional generalisations of the telegrapher's equation were considered in~\cite{compte,prw1,TK,polito,awad,gorska}, while non-Markovian discrete-time versions of the telegraph process were studied in~\cite{polito2}. For more details on the Cattaneo equation, as well as derivation and application of the telegrapher's equation, we refer to the literature, see, e.g.,~\cite{weiss2002,joseph,spigler,masoliverentropy}.

In the telegrapher's equation~(\ref{classical cattaneo eq}) it is assumed that the diffusion coefficient $D$ and the time interval $\tau$ are constants. In present paper we consider the case of space-dependent diffusion coefficient. In pure diffusion models, a space-dependent diffusivity is introduced to describe heterogeneous diffusion process (HDP), i.e., relative diffusion of passive tracers in the atmosphere~\cite{richardson,yaglom}, transport processes in heterogeneous media \cite{denisov1,Haggerty,Dentz,8,jstat,diss,chchme,ch1,ch3,ch2,me2020,edgarpre}, and on random fractals \cite{Procaccia}, including comb structures \cite{6,csf}. The mean first passage time and related search problems \cite{9,santos}, ergodicity breaking \cite{chchme,ch1,ch3} and infinite ergodic theory for HDPs~\cite{barkaiarxiv}, as well as L\'evy flights in inhomogeneous media \cite{10} have been investigated, as well.

In section~2 we introduce the telegrapher's equation for a finite-velocity HDP. We derive a general solution of the problem in section~3. Exact results for the probability density function (PDF) are obtained, and the asymptotic behaviours are analysed. In section~4 we present general result for the MSD, for which we observe different crossovers between diffusion regimes in the system. Several special cases are recovered, as well. We then introduce exponential resetting in the heterogeneous telegrapher's equation in section~5 and report exact results for the PDF and MSD. It is shown that in the long time limit the system approaches a non-equilibrium stationary state (NESS). The transition to the NESS is analysed in terms of the large deviation function. In section~6 we summarise our findings.

\section{From the Langevin equation with dichotomic noise to the telegrapher's equation}

The master equation for a persistent random walk leading to a Langevin equation with dichotomic noise was
considered in \cite{kac,balakrishnan}. The Langevin equation takes the form
\begin{align} \label{langevin homo}
    \dot{x}(t)=v\,\zeta(t),
\end{align}
where $v$ is a positive constant with physical dimension of a speed, and $\zeta(t)$ is a stationary dichotomic Markov process that jumps between two states $\pm1$ with the mean rate $\nu$, i.e., the inverse mean sojourn time for each state. The corresponding equation for the PDF $P(x,t)$ of such a process is the telegrapher's equation~(\ref{classical cattaneo eq}), where $\tau=\frac{1}{2\nu}$ and $D=v^2\tau$. For an elegant and simple derivation of equation~(\ref{classical cattaneo eq}) starting from equation~(\ref{langevin homo}) we refer the reader to Ref. \cite{balakrishnan}. In the limit $\nu\rightarrow\infty$, $v\rightarrow\infty$ such that $D$ is a finite constant the dichotomic noise reduces to a Gaussian white noise, and the diffusion equation is obtained. 
%This result was obtained by Marc Kac in 1974~\cite{kac}.
%{\color{blue}{For didactic purposes we repeat the basic steps of the derivation of telegrapher's equation. Kac considered a two-state system, such that a particle moving with a constant speed reverse its direction after a collision. The number of collisions experienced by the particle up to time $t$ is a Poisson process. The particle moving in one direction with velocity $v$ after the collision it changes its direction to $-v$, and then to $v$ after the next collision, etc. Since the number of collisions up to time $t$ of the homogeneous Poisson process with rate $\nu$ is $N(t)$, the velocity of the particle is given by
%$$v(t)=v\,(-1)^{N(t)}.$$So, the velocity takes the values $\pm v$ with equal probabilities $1/2$. The particle displacement then is $$x(t)=\int_{0}^{t}v(\tau)\,d\tau=v\int_{0}^{t}(-1)^{N(\tau)}\,d\tau.$$From here, one can find that the PDF satisfies the telegrapher's equation~(\ref{classical cattaneo eq}), where $\tau=\frac{1}{2\nu}$ and $D=v^2\tau$.}}

There exist different generalisations of the standard telegrapher's equation for inhomogeneous cases~\cite{weiss2002,masoliver1994,masoliverweiss,brissaud,kitahara,sancho0,ratanov99}. In this paper, we consider the form originating from the general nonlinear Langevin equation with multiplicative dichotomic noise
%~\cite{brissaud,kitahara,sancho0,ratanov99},
\begin{align}\label{le_ratanov}
    \dot{x}(t)=v(x)\zeta(t),
\end{align}
where $v(x)>0$ is a position-dependent speed, and $\zeta(t)$ is the same dichotomic process as in equation ~(\ref{langevin homo}). The result is
%again a stationary dichotomic Markov process that jumps between two values $\pm1$ with a mean rate $\nu$, such that $\tau=\frac{1}{2\nu}$. This equation can be obtained by the same procedure as for the standard telegrapher's equation, by considering a process $x(t)$ which satisfies $$x(t)=\int_{0}^{t}(-1)^{N(\tau)}v(x(\tau))\,d\tau.$$Again, $N(t)$ is the number of events up to time $t$ of a homogeneous Poisson process with rate $\nu$, and the position-dependent velocity takes the values $\pm v(x)$ with equal probabilities, where $v(x)$ is some positive continuous function~\cite{ratanov99}}}. The corresponding equation for the PDF $P(x,t)$ takes the form of heterogeneous telegrapher's equation
\begin{align}\label{te_ratanov}
    \frac{\partial^2}{\partial t^2}P(x,t)+\frac{1}{\tau}\frac{\partial}{\partial t}P(x,t)=\frac{\partial}{\partial x}\left\{v(x)\frac{\partial}{\partial x}\left[v(x)P(x,t)\right]\right\},
\end{align}
where $v(x)=\sqrt{D(x)/\tau}$. A detailed derivation of the heterogeneous telegrapher's equation from the Langevin equation~(\ref{le_ratanov}) is given in Ref.~\cite{ratanov99}, see Theorem~3.1.
%Detailed derivation of Eq.~(\ref{te_ratanov}) from Eq.~(\ref{le_ratanov}) is given in Ref.~\cite{ratanov99}, see Theorem~3.1}}. Here we note that the telegrapher's equation can not be obtained from stochastic differential equation with white Gaussian noise. Instead, for that we need the stochastic differential equation with dichotomic noise, which implies that there are no different interpretations for stochastic integrals as in case of HDP, and %thus, the It\^o-Stratonovich dilemma and its generalization do not exist. 
Other forms of the heterogeneous telegrapher's equation can be derived for a voltage and current in a inhomogeneous lossy transmission line, from the generalised Fick's law with position-dependent diffusivity, as well as from the persistent random walk in inhomogeneous medium. For details please see Appendix~\ref{app_different forms}. The motivation to consider heterogeneous models comes from the application of the telegrapher's equation in the description of turbulent diffusion~\cite{turb, turb1,turb2}, as well as in cosmic-ray transport~\cite{cosmic}. Moreover, the heterogeneous telegrapher's equation may also be important in the description of turbulent relative dispersion of particle pairs \cite{turbulent1,turbulent2} and represents a generalization of the Richardson model \cite{richardson}, since it takes into consideration the long-time correlation of the Lagrangian relative velocity of a particle pair, which exists in turbulent flows~\cite{sawford,sokolov}.

\section{Solution of the heterogeneous telegrapher's equation}

\subsection{Solution for $x_0\neq0$}

%Introducing the parameters $\tau=\frac{1}{2\nu}$ and $\mathcal{D}_{\alpha}=\frac{c^2}{2\nu} $, 
In what follows we consider equation of form (\ref{te_ratanov}) for power-law position dependent speed $v(x)=v_{\alpha}|x|^{\frac{\alpha}{2}}$, where $v_{\alpha}>0$ is with physical dimension $[v_{\alpha}]=\text{m}^{1-\alpha/2}s^{-1}$. Thus, we write the heterogeneous telegrapher's equation~(\ref{te_ratanov}) in the form
\begin{align}\label{cattaneo eq heterogeneous} 
\tau \frac{\partial^{2}}{\partial t^{2}}P(x,t)+\frac{\partial}{\partial t}P(x,t)= {\mathcal{D}_{\alpha}}\frac{\partial}{\partial x}\left\{|x|^{\frac{\alpha}{2}}\frac{\partial}{\partial x}\left[|x|^{\frac{\alpha}{2}}P(x,t)\right]\right\},
\end{align}
where $\mathcal{D}_{\alpha} = v^2_{\alpha} \tau$ is a diffusion coefficient with physical dimension $[\mathcal{D}_{\alpha}]=\text{m}^{2-\alpha}s^{-1}$. For $\tau\rightarrow 0$ and $v_\alpha\rightarrow\infty$ such that $\mathcal{D}_{\alpha} = const$, equation~(\ref{cattaneo eq heterogeneous}) becomes the heterogeneous (infinite-velocity) diffusion equation~\cite{chchme},\footnote{Here we note that we use $p(x,t)$ for the PDF of the HDP, while $P(x,t)$ is the PDF of the heterogeneous telegrapher's process.}
\begin{align}\label{diffusion eq heterogeneous} \frac{\partial}{\partial t}p(x,t)=\mathcal{D}_{\alpha}\frac{\partial}{\partial x}\left\{|x|^{\frac{\alpha}{2}}\frac{\partial}{\partial x}\left[|x|^{\frac{\alpha}{2}}p(x,t)\right]\right\},
\end{align}
which is derived from the Langevin equation in the Stratonovich interpretation~\cite{chchme}
\begin{align}\label{LE}
\dot{x}(t)=\sqrt{2\,\mathcal{D}_{\alpha}|x|^{\alpha}}\,\eta(t),    
\end{align}
with position dependent diffusion coefficient, where $\eta(t)$ is a white Gaussian noise of zero mean. Here we use $\alpha<2$ to ensure the growth condition for existence and uniqueness of the solution of a Markovian stochastic differential equation, see Ref.~\cite{chchme}. The case with $\alpha=2$ requires a separate consideration and is related to the problem of geometric Brownian motion in the Stratonovich interpretation~\cite{pre2020}.

To solve equation~(\ref{cattaneo eq heterogeneous}), we consider the initial conditions\footnote{See the discussion of initial conditions in Refs.~\cite{weiss2002,ejp,ratanovbook}.} \begin{align}\label{initial condition}
P(x,t=0)=\delta(x-x_0), \quad \frac{\partial}{\partial t}P(x,t=0)=0,
\end{align}
and the boundary conditions are set to zero at infinity, i.e., 
\begin{align}
    P(\pm\infty,t)=0, \quad \frac{\partial}{\partial x}P(\pm\infty,t)=0.\nonumber
\end{align} 
The Laplace transform\footnote{The Laplace transform is defined by $\hat{f}(s)=\mathcal{L}\left[f(t)\right]=\int_{0}^{\infty}f(t)\,e^{-st}dt$, while the inverse Laplace transform by $f(t)=\mathcal{L}^{-1}\left[\hat{f}(s)\right]=\frac{1}{2\pi\imath}\int_{c-\imath\infty}^{c+\imath\infty}\hat{f}(s)\,e^{s t}ds$.} of equation~(\ref{cattaneo eq heterogeneous}) yields
\begin{align}\label{cattaneo eq_general laplace} 
s(1+\tau s)\hat{P}(x,s)-(1+\tau s)\delta(x-x_0) =\mathcal{D}_{\alpha}\frac{\partial}{\partial x}\left\{|x|^{\frac{\alpha}{2}}\frac{\partial}{\partial x}\left[|x|^{\frac{\alpha}{2}}\hat{P}(x,s)\right]\right\},
\end{align}
which can be rewritten in the form
\begin{align}\label{cattaneo eq_general laplace2} 
s\hat{P}(x,s)-\delta(x-x_0) =\mathcal{D}_{\alpha}\frac{1}{1+\tau s}\frac{\partial}{\partial x}\left\{|x|^{\frac{\alpha}{2}}\frac{\partial}{\partial x}\left[|x|^{\frac{\alpha}{2}}\hat{P}(x,s)\right]\right\}.
\end{align}
We note that by inverse Laplace transform, we obtain an equivalent formulation for equation~(\ref{cattaneo eq heterogeneous}),
\begin{align}\label{diffusion eq heterogeneous final} \frac{\partial}{\partial t}P(x,t)=\mathcal{D}_{\alpha}\int_{0}^{t}K(t-t',\tau)\frac{\partial}{\partial x}\left\{|x|^{\frac{\alpha}{2}}\frac{\partial}{\partial x}\left[|x|^{\frac{\alpha}{2}}P(x,t')\right]\right\}dt',
\end{align}
where $$K(t,\tau)=\frac{1}{\tau}e^{-t/\tau} \quad \rightarrow \quad \hat{K}(s,\tau)=\frac{1}{1+s\,\tau}.$$Therefore, the heterogeneous telegrapher's equation~(\ref{cattaneo eq heterogeneous}) can be considered as a heterogeneous diffusion equation with an non-local memory kernel. Such an equation with exponential memory kernel was analysed for $\alpha=0$ in~\cite{sokolov2002,chechkinpre2021}.

\begin{figure}
\centering{(a)\includegraphics[width=8cm]{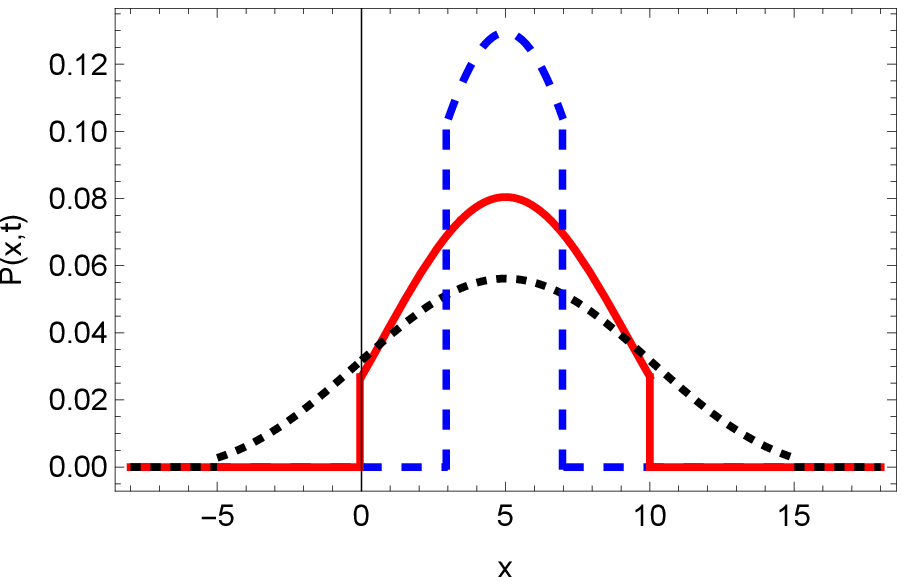} (b)\includegraphics[width=8cm]{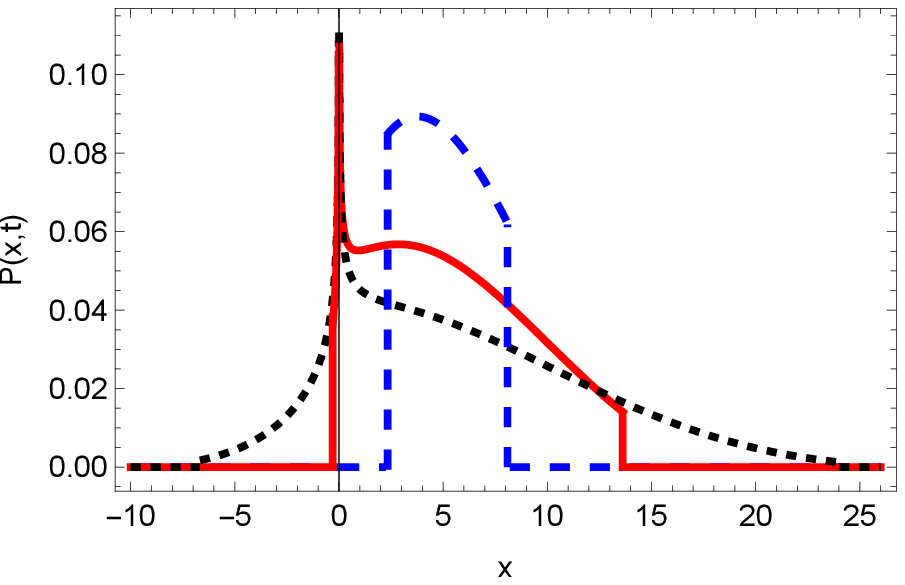}\\
\medskip 
(c)\includegraphics[width=8cm]{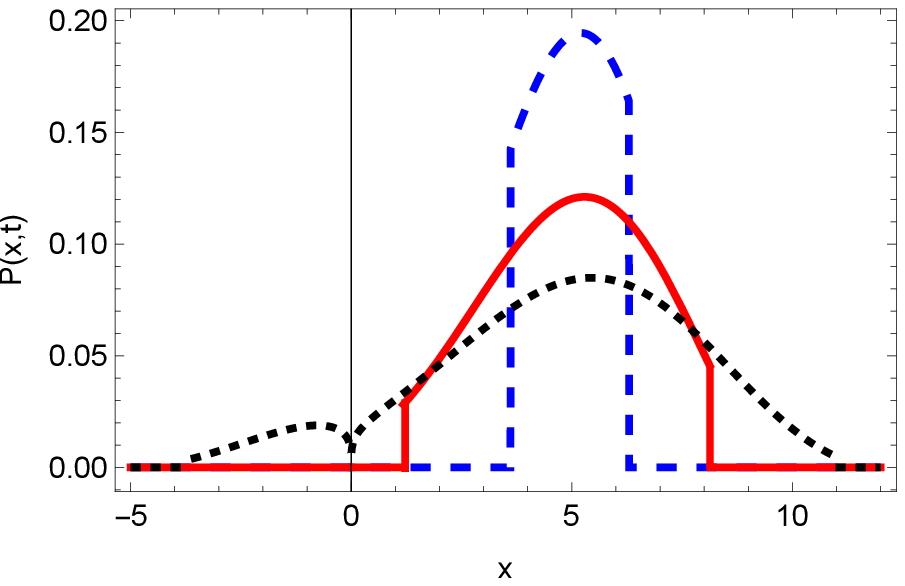} (d)\includegraphics[width=8cm]{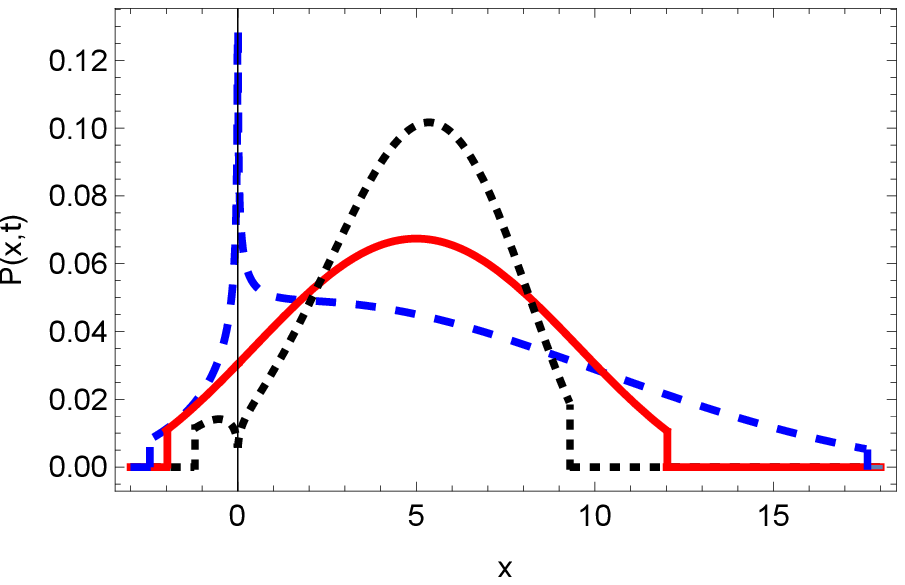}
\caption{PDF~(\ref{final pdf}) for $x_0=5$, $\tau=1$, $v_\alpha=1$. (a) $\alpha=0$, $t=2$ (blue dashed line), $t=5$ (red solid line), $t=10$ (black dotted line). (b) $\alpha=0.5$, $t=2$ (blue dashed line), $t=5$ (red solid line), $t=10$ (black dotted line). (c) $\alpha=-0.5$, $t=2$ (blue dashed line), $t=5$ (red solid line), $t=10$ (black dotted line). (d) Comparison between PDFs for $t=7$, and $\alpha=0.5$ (blue dashed line), $\alpha=0$ (red solid line), $\alpha=-0.5$ (black dotted line). The delta functions at
the endpoints, which ensure normalization of the PDFs are not shown in the figure.}\label{fig_pdf_te_hdp}}
\end{figure}

\begin{figure}[t!]
\centering{(a)\includegraphics[width=8cm]{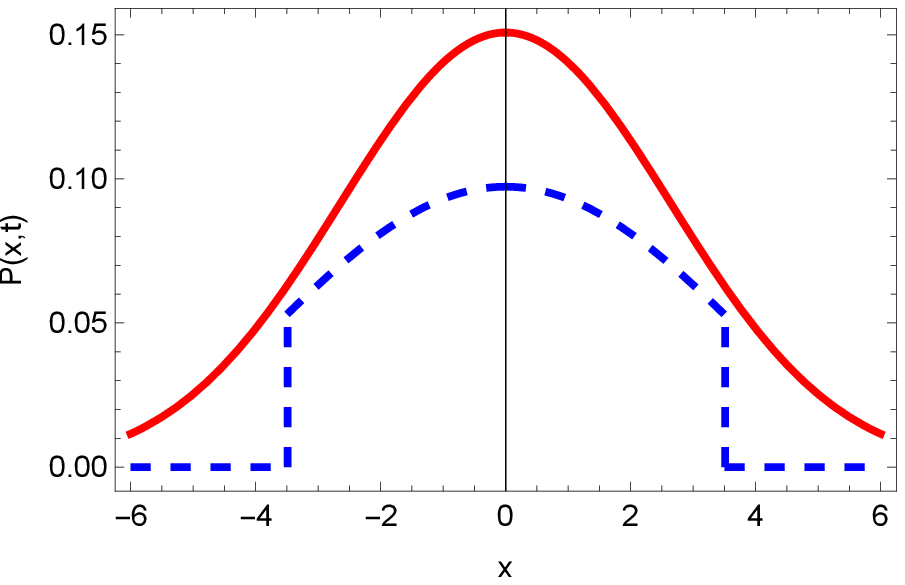} \\
\medskip 
(b)\includegraphics[width=8cm]{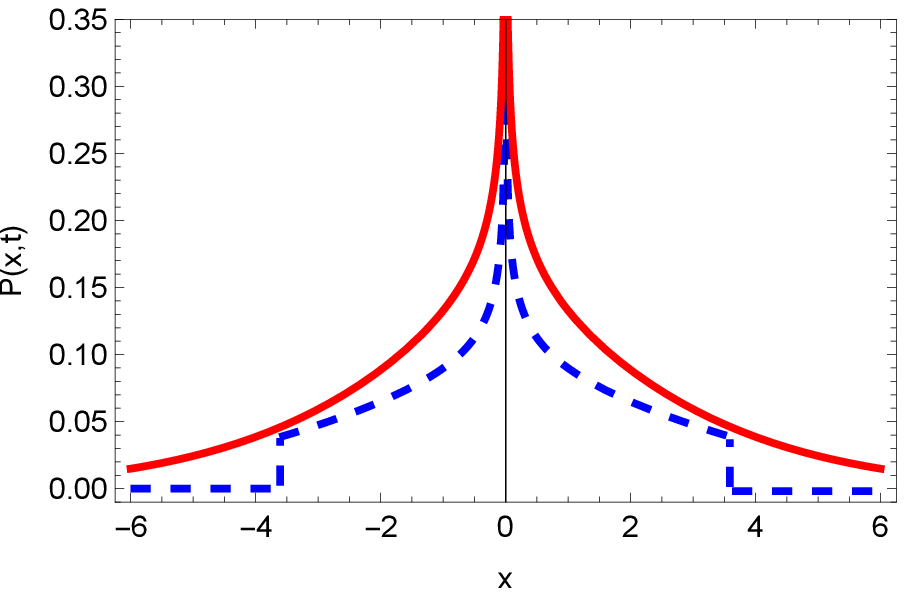} (c)\includegraphics[width=8cm]{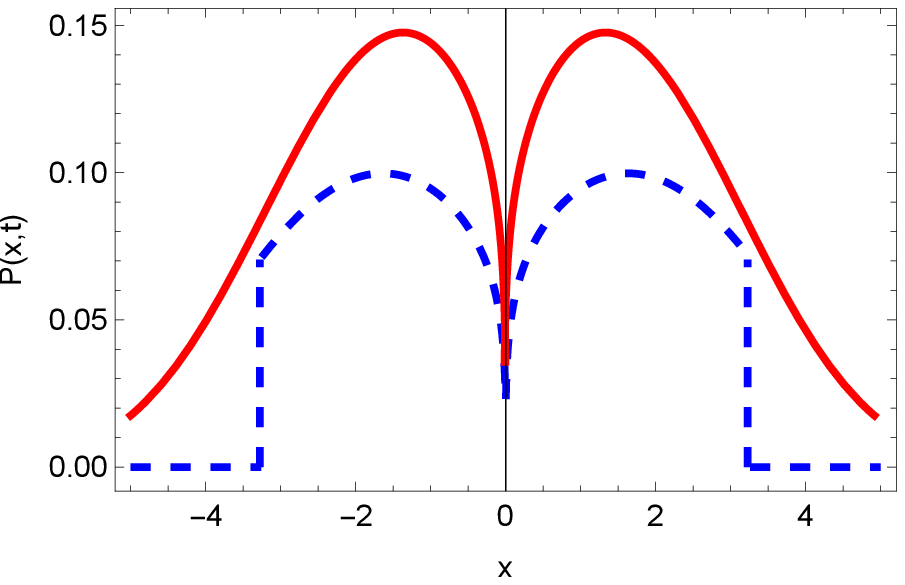}
\caption{PDF (\ref{final pdf x0=0}) for $\tau=1$, $v_\alpha=1$ (blue dashed line) versus PDF (\ref{pdf hdp diffusion}) (red solid line) for $\mathcal{D}_\alpha=1$, at $t=3.5$. (a) $\alpha=0$ -- standard telegrapher's process versus standard diffusion; (b) $\alpha=0.5$ (long-time superdiffusion); (c) $\alpha=-0.5$ (long-time subdiffusion). The contribution of the delta function at the endpoints in the solution of the telegrapher's equation (blue dashed lines), which ensure normalization of the PDF, are not shown in the figure.}\label{fig_pdf_te_vs_diff_hdp}}
\end{figure}

For $\tau=0$, we obtain the HDP equation in Laplace space,
\begin{align}\label{diffusion eq_general laplace2} 
s\hat{p}(x,s)-\delta(x-x_0) =\mathcal{D}_{\alpha}\frac{\partial}{\partial x}\left\{|x|^{\frac{\alpha}{2}}\frac{\partial}{\partial x}\left[|x|^{\frac{\alpha}{2}}\hat{p}(x,s)\right]\right\}.
\end{align}
Using $s\rightarrow s(1+\tau s)$ in equation (\ref{diffusion eq_general laplace2}) we obtain
\begin{align}\label{diffusion eq_general laplace2_2} 
s(1+\tau s)\hat{p}(x,s(1+\tau s))-\delta(x-x_0) =\mathcal{D}_{\alpha}\frac{\partial}{\partial x}\left\{|x|^{\frac{\alpha}{2}}\frac{\partial}{\partial x}\left[|x|^{\frac{\alpha}{2}}\hat{p}(x,s(1+\tau s))\right]\right\}.
\end{align}
Now, one can see that by substituting
\begin{align}\label{P vs p}
    \hat{P}(x,s)=(1+s\,\tau)\,\hat{p}(x,s(1+s\,\tau)),
\end{align}
from equation~(\ref{diffusion eq_general laplace2_2}) we arrive at equation~(\ref{cattaneo eq_general laplace2}). Thus, we can directly obtain the solution $P(x,t)$ from the solution $p(x,t)$. Relation (\ref{P vs p}) is not affected by the inhomogeneity in space and was derived in~\cite{sokolov2002} for the standard telegrapher's equation, Eq.~(\ref{diffusion eq heterogeneous final}) with $\alpha=0$, by using the subordination approach.
 %where it is shown that $\hat{P}(x,s)=\frac{1}{\hat{K}(s,\tau)}\hat{p}\left(x,\frac{s}{\hat{K}(s,\tau)}\right)$, see Eq.~(11) in~\cite{sokolov2002}.}} %{\color{blue}{Such procedure has been used within the subordination approach, where the PDF of the subordinated process can be derived from the PDF of the parent process, see for example~\cite{marcin1,marcin2}.}} 
The solution of the diffusion equation for the HDP (\ref{diffusion eq heterogeneous}) is given by (see equation~(18) in~\cite{hdpjpa2022}) \begin{align}\label{pdf hdp diffusion}
    p(x,t)=\frac{|x|^{1/p-1}}{\sqrt{4\pi\mathcal{D}_{p}t}}\times \exp\left(-\frac{p^2\left|\text{sgn}(x)|x|^{1/p}-\text{sgn}(x_0)|x_0|^{1/p}\right|^2}{4\mathcal{D}_p t}\right),
\end{align}
where $\mathcal{D}_{\alpha}\rightarrow\mathcal{D}_p$ and $p=\frac{2}{2-\alpha}$, and which in Laplace space reads
\begin{align}\label{p solution laplace}
    \hat{p}(x,s)=\frac{|x|^{1/p-1}}{2\sqrt{\mathcal{D}_p}}s^{-1/2}\times \exp\left(-\frac{p}{\sqrt{\mathcal{D}_p}}\left|\text{sgn}(x)|x|^{1/p}-\text{sgn}(x_0)|x_0|^{1/p}\right|s^{1/2}\right).
\end{align}
By integration of (\ref{p solution laplace}) it is shown that $\int_{-\infty}^{\infty}\hat{p}(x,s)\,dx=\frac{1}{s}$ (see Appendix~A in Ref.~\cite{hdpjpa2022}), which means that $p(x,t)$ is normalized. Therefore, from Eq.~(\ref{P vs p}), we have
\begin{align}\label{normalization P}
    \int_{-\infty}^{\infty}\hat{P}(x,s)\,dx&=(1+s\,\tau)\int_{-\infty}^{\infty}\hat{p}(x,s(1+s\,\tau))\,dx=(1+s\,\tau)\frac{1}{s(1+s\,\tau)}=\frac{1}{s},
\end{align}
which means that the PDF $P(x,t)$ is normalized, as well. We also note that in the limit $\alpha\rightarrow2$ and $x, x_0 > 0$, the PDF (18) turns into the log-normal distribution for geometric Brownian motion~\cite{pre2020}. Thus, for the PDF $P(x,t)$, we obtain in Laplace space
\begin{align}\label{pdf laplace final}
    \hat{P}(x,s)=&\frac{|x|^{1/p-1}}{2v_p}(\tau^{-1}+s)(s+\tau^{-1})^{-1/2}s^{-1/2} \nonumber\\&\times \exp\left(-\frac{p}{v_p}\left|\text{sgn}(x)|x|^{1/p}-\text{sgn}(x_0)|x_0|^{1/p}\right|(s+\tau^{-1})^{1/2}s^{1/2}\right),
\end{align}
where $v_p=\sqrt{\mathcal{D}_p/\tau}$. This result can be rewritten in the form
\begin{align}
    \hat{P}(x,s)=\frac{|x|^{1/p-1}}{2v_p}(\tau^{-1}+s)\hat{L}(x,s),
\end{align}
with
\begin{align}
    \hat{L}(x,s)=&(s+\tau^{-1})^{-1/2}s^{-1/2}\nonumber\\&\times \exp\left(-\frac{p}{v_p}\left|\text{sgn}(x)|x|^{1/p}-\text{sgn}(x_0)|x_0|^{1/p}\right|(s+\tau^{-1})^{1/2}s^{1/2}\right).
\end{align}
The PDF then becomes
\begin{align}\label{P through L}
    P(x,t)=\frac{|x|^{1/p-1}}{2v_p}\left[\tau^{-1}L(x,t)+\frac{\partial}{\partial t}L(x,t)\right],
\end{align}
where
\begin{align}
    L(x,t)=&\theta\left(v_p t-p\left|\text{sgn}(x)|x|^{1/p}-\text{sgn}(x_0)|x_0|^{1/p}\right|\right)
    e^{-\frac{t}{2\tau}} \nonumber\\& \times I_{0}\left(\frac{\sqrt{v_p^2t^2-p^2\left[\text{sgn}(x)|x|^{1/p}-\text{sgn}(x_0)|x_0|^{1/p}\right]^2}}{2v_p \tau}\right).
\end{align}
Finally, we obtain the form
\begin{align}\label{final pdf}
    P(x,t)
    =&\frac{|x|^{1/p-1}}{2}e^{-\frac{t}{2\tau}}\delta\left(v_p t-p\left|\text{sgn}(x)|x|^{1/p}-\text{sgn}(x_0)|x_0|^{1/p}\right|\right)\nonumber\\&+\frac{|x|^{1/p-1}}{4v_p\tau}\theta\left(v_p t-p\left|\text{sgn}(x)|x|^{1/p}-\text{sgn}(x_0)|x_0|^{1/p}\right|\right)e^{-\frac{t}{2\tau}}\left[I_{0}(\xi)+\frac{t}{2\tau}\frac{I_1(\xi)}{\xi}\right],
\end{align}
where 
\begin{align}
    \xi=\frac{\sqrt{t^2-\frac{p^2}{v_p^2}\left[\text{sgn}(x)|x|^{1/p}-\text{sgn}(x_0)|x_0|^{1/p}\right]^2}}{2\tau}.
\end{align}
Here we note that $I_{\nu}(z)$ is the modified Bessel function of the first kind with asymptotic behavior $I_{\nu}(z)\sim\frac{e^{z}}{\sqrt{2\pi z}}$. Thus, in the long time limit, from (\ref{final pdf}) we arrive at the PDF (\ref{pdf hdp diffusion}) for the heterogeneous diffusion equation, while in short time limit we have
\begin{align}\label{final pdf_diffusion}
    P(x,t)=\frac{|x|^{1/p-1}}{2}\delta\left(v_p t-p\left|\text{sgn}(x)|x|^{1/p}-\text{sgn}(x_0)|x_0|^{1/p}\right|\right).
\end{align}
For $\alpha=0$, i.e., $p=1$, we obtain the solution for the standard telegrapher's equation~\cite{goldstein} (see also equation~(16) in Ref.~\cite{masporraweiss}, equation~(2.5.3) in Ref.~\cite{ratanovbook} or equation~(6) in \cite{masoliverreset}) 
\begin{align}\label{solution standard TE}
P(x,t)=&\frac{e^{-\frac{t}{2\tau}}\delta\left(vt-|x-x_0|\right)}{2}\nonumber\\&+\frac{e^{-\frac{t}{2\tau}}}{4v\tau}\theta\left(vt-|x-x_0|\right)\left[I_{0}\left(\frac{\sqrt{v^2t^2-|x-x_0|^{2}}}{2v\tau}\right)+vt\frac{I_1\left(\frac{\sqrt{v^2t^2-|x-x_0|^{2}}}{2v\tau}\right)}{\sqrt{v^2t^2-|x-x_0|^{2}}}\right].
\end{align}
For $x_0=0$, the PDF~(\ref{final pdf}) reduces to 
\begin{align}\label{final pdf x0=0}
    P(x,t)
    =\frac{|x|^{1/p-1}}{2}e^{-\frac{t}{2\tau}}\delta\left(v_p t-p|x|^{1/p}\right)+&\frac{|x|^{1/p-1}}{4v_p\tau}e^{-\frac{t}{2\tau}}\theta\left(v_p t-p|x|^{1/p}\right)\nonumber\\&\times\left[I_{0}\left(\frac{\sqrt{v_p^2t^2-p^2|x|^{2/p}}}{2v_p\tau}\right)+v_p t\frac{I_1\left(\frac{\sqrt{v_p^2t^2-p^2|x|^{2/p}}}{2v_p\tau}\right)}{\sqrt{v_p^2t^2-p^2|x|^{2/p}}}\right].
\end{align}
A graphical representation of the PDF~(\ref{final pdf}) for different parameter values is shown in figure~\ref{fig_pdf_te_hdp}. We observe the distinct finite-velocity propagation where the PDF drops to zero beyond the front $v_p t=p\left|\text{sgn}(x)|x|^{1/p}-\text{sgn}(x_0)|x_0|^{1/p}\right|$. We compare the PDFs for the telegrapher's equation and the diffusion equation in figure~\ref{fig_pdf_te_vs_diff_hdp}. Due to the finite-velocity propagation, we see the drop to zero of the PDF for the telegrapher's process (red solid lines), while instantaneous propagation is observed for the HDP (blue dashed lines). We also note that the PDF is unimodal for $\alpha>0$ and bimodal for $\alpha<0$. Here we note that the contribution of the delta functions which ensure normalization of the PDF at
the endpoints $x$, that satisfy $v_p t=p\left|\text{sgn}(x)|x|^{1/p}-\text{sgn}(x_0)|x_0|^{1/p}\right|$, are not shown in the figures. We have already proven in Eq.~(\ref{normalization P}) that the PDF satisfies the normalization condition $\int_{-\infty}^{\infty}P(x,t)\,dx=1$, however, for the readers' convenience, in Appendix~\ref{app norm} we provide additional proof of the normalization directly by integration of the PDF~(\ref{P through L}).

\subsection{Alternative solution for $x_0=0$}

For the specific initial condition at the origin, $P(x,0)=\delta(x)$, we find the solution of equation~(\ref{cattaneo eq heterogeneous}) in Laplace space yielding
\begin{eqnarray}\label{cattaneo eq_general laplace 3} 
s(1+\tau s)\hat{P}(x,s)-(1+\tau s)\delta(x) =\mathcal{D}_{\alpha}\frac{\partial}{\partial x}\left\{|x|^{\frac{\alpha}{2}}\frac{\partial}{\partial x}\left[|x|^{\frac{\alpha}{2}}\hat{P}(x,s)\right]\right\}.
\end{eqnarray}
From here, by differentiation with respect to $x$, we find
\begin{align}\label{cattaneo eq heterogeneous_general laplace2}
s(1+\tau s)\hat{P}(x,s)-(1+\tau s)\delta(x)=\mathcal{D}_{\alpha}\frac{\partial}{\partial x}\left\{(2\theta(x)-1)\frac{\alpha}{2}|x|^{\alpha-1}\hat{P}(x,s)+|x|^{\alpha}\frac{\partial}{\partial x}\hat{P}(x,s)\right\},
\end{align}
where $\theta(x)$ is the Heaviside step function. By further differentiation with respect to $x$, we obtain
\begin{align}\label{cattaneo eq heterogeneous_general laplace3}
s(1+\tau s)\hat{P}(x,s)-(1+\tau s)\delta(x)=&\mathcal{D}_{\alpha}\left\{\frac{\alpha(\alpha-1)}{2}|x|^{\alpha-2}\hat{P}(x,s)+\alpha|x|^{\alpha-1}\delta(x)\hat{P}(x,s)\right.\nonumber\\&\left.+\frac{3\alpha}{2}\left[2\theta(x)-1\right]|x|^{\alpha-1}\frac{\partial}{\partial x}\hat{P}(x,s)+|x|^{\alpha}\frac{\partial^2}{\partial x^2}\hat{P}(x,s)\right\}.
\end{align}
We see that equation~(\ref{cattaneo eq heterogeneous_general laplace3}) is invariant with respect to inversion $x\rightarrow  -x$, and we use $y=|x|$. Equation~(\ref{cattaneo eq heterogeneous_general laplace3}), then becomes
\begin{align}\label{cattaneo eq heterogeneous2 laplace4} 
s(1+\tau s)\hat{P}(y,s)-(1+\tau s)\delta(x)=&\mathcal{D}_{\alpha}\frac{(\alpha-1)\alpha}{2}y^{\alpha-2}\hat{P}(y,s)+\mathcal{D})_{\alpha}\alpha y^{\alpha-1}\hat{P}(y,s)\delta(x) \nonumber\\&+\mathcal{D}_{\alpha}\frac{3\alpha}{2}y^{\alpha-1}\frac{\partial}{\partial y}\hat{P}(y,s)+2\mathcal{D}_{\alpha}y^{\alpha}\frac{\partial}{\partial y}\hat{P}(y,s)\delta(x)+\mathcal{D}_{\alpha}y^{\alpha}\frac{\partial^2}{\partial y^2}\hat{P}(y,s),
\end{align}
where $\hat{P}(|x|,s)=\mathcal{C}(s)\hat{f}(|x|,s)=\mathcal{C}(s)\hat{f}(y,s)$, and $\mathcal{C}(s)$ is a function of $s$. From equation~(\ref{cattaneo eq heterogeneous2 laplace4}), we obtain a system of two equations
\begin{align}\label{system1}
\frac{\partial^2}{\partial y^2}\hat{f}(y,s)+\frac{3\alpha/2}{y}\frac{\partial}{\partial y}\hat{f}(y,s)+\left[-\frac{s(1+s\tau)}{\mathcal{D}_{\alpha}}y^{-\alpha}+\frac{(\alpha-1)\alpha}{2}\frac{1}{y^2}\right]\hat{f}(y,s)=0,
\end{align}
\begin{align}\label{system2}
-(1+s\tau)=\mathcal{C}(s)\,\mathcal{D}_{\alpha}\left.\left[\alpha y^{\alpha-1}\hat{f}(y,s)+2y^{\alpha}\frac{\partial}{\partial y}\hat{f}(y,s)\right]\right|_{y=0}.
\end{align}
Equation~(\ref{system1}) is the Lommel-type equation
\begin{align}\label{lommel}
z''(y)+\frac{1-2\beta'}{y}z'(y)+\left[\left(a\alpha'y^{\alpha'-1}\right)^{2}+\frac{\beta'^{2}-\nu^{2}\alpha'^{2}}{y^2}\right]z(y)=0,
\end{align}
which solution is given by $$z(y)=y^{\beta'}Z_{\nu}\left(ay^{\alpha'}\right),$$ where $Z_{\nu}(y)=C_{1}J_{\nu}(y)+C_{2}N_{\nu}(y)$ and $J_{\nu}(y)$ and $N_{\nu}(y)$ are the Bessel functions of first and second kind, respectivelly. Therefore, we have
\begin{align}\label{sol1}
\hat{f}(y,s)=y^{\frac{2-3\alpha}{4}}\,Z_{\frac{1}{2}}\left(\imath\frac{2}{2-\alpha}\sqrt{\frac{s(1+s\tau)}{\mathcal{D}_{\alpha}}}y^{\frac{2-\alpha}{2}}\right)=y^{\frac{2-3\alpha}{4}}\,K_{\frac{1}{2}}\left(\frac{2}{2-\alpha}\sqrt{\frac{s(1+s\tau)}{\mathcal{D}_{\alpha}}}y^{\frac{2-\alpha}{2}}\right),
\end{align} 
where $Z_{\nu}(\imath z)=C_{1}I_{\nu}(z)+C_{2}K_{\nu}(z)$, and $I_{\nu}(z)$ and $K_{\nu}(y)$ are the modified Bessel functions~\cite{bookGR}, where $K_{\nu}(z)$ satisfies the zero boundary conditions at infinity. The PDF then reads
\begin{align}\label{sol1 pdf}
\hat{P}(x,s)=\mathcal{C}(s)\,|x|^{\frac{2-3\alpha}{4}}\,K_{\frac{1}{2}}\left(\frac{2}{2-\alpha}\sqrt{\frac{s(1+s\tau)}{\mathcal{D}_{\alpha}}}|x|^{\frac{2-\alpha}{2}}\right).
\end{align} 
From equation~(\ref{system2}), by using the series representation of $K_{\nu}(y)$
\begin{align}\label{K series}
K_{\nu}(z)\sim\frac{\Gamma(\nu)}{2}\left(\frac{z}{2}\right)^{-\nu}\left[1+\frac{z^{2}}{4(1-\nu)}+\dots\right]+\frac{\Gamma(-\nu)}{2}\left(\frac{z}{2}\right)^{\nu}\left[1+\frac{z^{2}}{4(\nu+1)}+\dots\right],
\end{align}
for $z\rightarrow0$ and $\nu\notin Z$, for $\mathcal{C}(s)$ we obtain
\begin{align}
     \mathcal{C}(s)=\mathcal{D}_{\alpha}^{-3/4}\,\frac{s^{-1/4}(1+s\tau)^{3/4}}{\sqrt{(2-\alpha)\pi}}.
\end{align}
The PDF in Laplace space then reads
\begin{align}\label{sol laplace s}
\hat{P}(x,s)&=\mathcal{D}_{\alpha}^{-3/4}\,\frac{s^{-1/4}(1+s\tau)^{3/4}}{\sqrt{(2-\alpha)\pi}}|x|^{\frac{2-3\alpha}{4}}K_{\frac{1}{2}}\left(\frac{2}{2-\alpha}\sqrt{\frac{s(1+s\tau)}{\mathcal{D}_{\alpha}}}|x|^{\frac{2-\alpha}{2}}\right)\nonumber\\&=\frac{|x|^{-\alpha/2}}{2v_{\alpha}}\left(s+\tau^{-1}\right)^{1/2}s^{-1/2}\exp\left(-\frac{2}{2-\alpha}\frac{s^{1/2}\left(s+\tau^{-1}\right)^{1/2}}{v_{\alpha}}|x|^{(2-\alpha)/2}\right),
\end{align}
where we use $K_{\frac{1}{2}}(x)=\sqrt{\pi/(2x)}\,e^{-x}$. This is the same result as (\ref{pdf laplace final}) for $x_0=0$, as it should be.

\section{Calculation of the mean squared displacement}

From the PDF~(\ref{P vs p}), we find the MSD $\left\langle \hat{x}^2(s)\right\rangle_\tau=\int_{-\infty}^{\infty}x^2\,\hat{P}(x,s)\,dx$,
\begin{align}\label{msd laplace te hdp}
    \left\langle \hat{x}^2(s)\right\rangle_\tau&=(1+s\,\tau)\int_{-\infty}^{\infty}x^2\,\hat{p}(x,s(1+s\,\tau))\,dx=(1+s\,\tau)\left\langle\hat{x}^2(s(1+s\tau))\right\rangle_{0},
\end{align}
where $\left\langle x^2(u)\right\rangle_{0}=\frac{\Gamma\left(1+2p\right)}{p^{2p}}\frac{(\mathcal{D}_p u)^{p}}{\Gamma\left(1+p\right)}\,{_1}F_{1}\left(-p,\frac{1}{2},-p^{2}\frac{|x_{0}|^{2/p}}{4\mathcal{D}_p u}\right)$ is the MSD for $\tau=0$, see equation~(20) in Ref.~\cite{hdpjpa2022}. Here $_1F_1(a,b,z)$ is the confluent hypergeometric function of the first kind. 

\begin{figure}[t!]
\centering{\includegraphics[width=8cm]{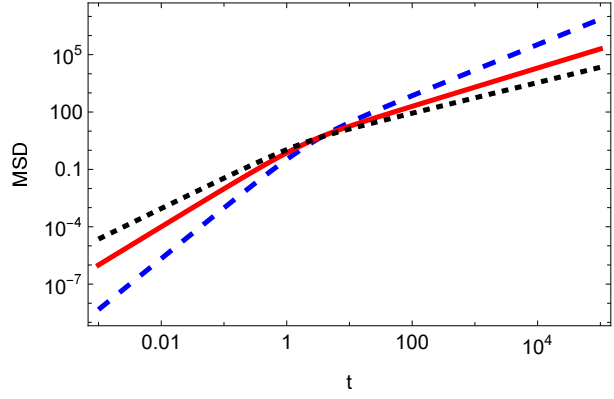}
\caption{MSD~(\ref{msd s}) for $\tau=1$, $\mathcal{D}_\alpha=1$, and $\alpha=0.5$ (blue dashed line), $\alpha=0$ (red solid line), $\alpha=-0.5$ (black dotted line).}\label{fig_msd}}
\end{figure}

For $x_0=0$, the MSD is $\left\langle x^2(u)\right\rangle_{0}=\frac{\Gamma\left(1+2p\right)}{p^{2p}}\frac{(\mathcal{D}_p u)^{p}}{\Gamma\left(1+p\right)}$, i.e., $\left\langle \hat{x}^2(s)\right\rangle_{0}=\frac{\Gamma\left(1+2p\right)}{p^{2p}}\mathcal{D}_p u^{p}s^{-p-1}$, and thus, from equation~(\ref{msd laplace te hdp}) we find
\begin{align}\label{msd s}
\left\langle x^{2}(t)\right\rangle_{\tau}
 &=\frac{\Gamma\left(1+2p\right)}{p^{2p}}\left(\frac{\mathcal{D}_{p}}{\tau}\right)^{p}\mathcal{L}^{-1}\left[\frac{s^{-p-1}}{(s+\tau^{-1})^{p}}\right]
=\frac{\Gamma\left(1+2p\right)\left(\mathcal{D}_{p}\tau\right)^{p}}{p^{2p}}\left(\frac{t}{\tau}\right)^{2p}E_{1,2p+1}^{p}\left(-\frac{t}{\tau}\right),
\end{align}
where 
\begin{align}\label{three parameter ml}
E_{\rho,\beta}^{\delta}(z)=\sum_{n=0}^{\infty}\frac{(\delta)_{n}}{\Gamma(\rho n+\beta)}\frac{z^{n}}{n!}
\end{align} 
is the three-parameter Mittag-Leffler function \cite{prabhakar}, and $(\delta)_{n}=\Gamma(\delta+n)/\Gamma(\delta)$ is the Pochhammer symbol. To perform the inverse Laplace transform in Eq.~(\ref{msd s}) we use the formula, see Eq.~(5.1.33) in Ref.~\cite{mainardi book}, %The Laplace transform of~(\ref{three parameter ml}) is given by
\begin{eqnarray}\label{Laplace ML}
\mathcal{L}^{-1}\left[\frac{s^{\rho\delta-\beta}}{(s^\rho+\lambda)^\delta}\right]=t^{\beta-1}E_{\rho,\beta}^{\delta}\left(-\lambda{t}^{\rho}\right), %\mathcal{L}\left[t^{\beta-1}E_{\rho,\beta}^{\delta}\left(-\lambda{t}^{\rho}\right)\right]=\frac{s^{\rho\delta-\beta}}{(s^\rho+\lambda)^\delta},
\end{eqnarray}
with $|\lambda/s^{\rho}|<1$ (note that in Eq.~~(\ref{msd s}) we have $\rho\rightarrow1$, $\delta\rightarrow p$, and $\rho\delta-\beta\rightarrow-p-1$, which means $\beta\rightarrow2p+1$). %{\color{blue}{which we used it to perform the inverse Laplace transform in Eq.~(\ref{msd s})}}. 
From the definition of the Mittag-Leffler function and the known formula \cite{garrappa}
\begin{align}\label{GML_formula}
E_{\rho,\beta}^{\delta}(-z)=\frac{z^{-\delta}}{\Gamma(\delta)}\sum_{n=0}^{\infty}\frac{\Gamma(\delta+n)}{\Gamma(\beta-\rho(\delta+n))}\frac{(-z)^{-n}}{n!},
\end{align} 
with $z>1$, and $0<\rho<2$, we find the asymptotic behavior of the MSD in the short and long time limits,
\begin{align}\label{msd asympt s}
\left\langle x^{2}(t)\right\rangle_{\tau}\sim\frac{\Gamma\left(1+2p\right)}{p^{2p}}\left(\mathcal{D}_{p}\tau\right)^{p}\left\{\begin{array}{l l l}
\frac{\left(t/\tau\right)^{2p}}{\Gamma\left(1+2p\right)}, \quad t/\tau\ll1,\\ \\
\frac{(t/\tau)^{p}}{\Gamma\left(1+p\right)}, \quad t/\tau\gg1.
\end{array}\right.
\end{align}
From these results we take the following conclusions: {\bf (i)} for $0<\alpha<2$ we have a crossover from hyperdiffusion $\left\langle x^{2}(t)\right\rangle\simeq t^{\mu_1}$, $\mu_{1}=\frac{4}{2-\alpha}$, $\mu_{1}>2$ to $\left\langle x^{2}(t)\right\rangle\simeq t^{\mu_2}$, $\mu_{2}=\frac{2}{2-\alpha}$, which means: (a) either superdiffusion for $0<\alpha<1$, (b) ballistic motion for $\alpha=1$, (c) or hyperdiffusion for $1<\alpha<2$; {\bf (ii)} for $\alpha=0$ we observe a crossover from ballistic motion $\left\langle x^{2}(t)\right\rangle\simeq t^{2}$, to normal diffusion $\left\langle x^{2}(t)\right\rangle\simeq t$; {\bf (iii)} for $-2<\alpha<0$ we have a crossover from superdiffusion $\left\langle x^{2}(t)\right\rangle\simeq t^{\mu_1}$, $\mu_{1}=\frac{4}{2-\alpha}$, $1<\mu_{1}<2$ to subdiffusion with $\left\langle x^{2}(t)\right\rangle\simeq t^{\mu_2}$, $\mu_{2}=\frac{2}{2-\alpha}$, $1/2<\mu_{2}<1$; {\bf (iv)} for $\alpha=-2$ we have a crossover from normal diffusion $\left\langle x^{2}(t)\right\rangle\simeq t^{\mu_1}$, $\mu_{1}=\frac{4}{2-\alpha}=1$ to subdiffusion with $\left\langle x^{2}(t)\right\rangle\simeq t^{\mu_2}$, $\mu_{2}=\frac{2}{2-\alpha}=\frac{1}{2}$;
(iv) for $\alpha<-2$ we obtain a crossover from subdiffusion $\left\langle x^{2}(t)\right\rangle\simeq t^{\mu_1}$, $\mu_{1}=\frac{4}{2-\alpha}$, $0<\mu_{1}<1$ to subdiffusion with $\left\langle x^{2}(t)\right\rangle\simeq t^{\mu_2}$, $\mu_{2}=\frac{2}{2-\alpha}$, $0<\mu_{2}<1/2$. Therefore, various diffusive crossovers are observed, rendering the considered model a suitable basis for the description of anomalous dynamics in complex systems. The obtained results are summarized in Table~\ref{tab:my_label}. In figure~\ref{fig_msd}, we show the graphical representation of the MSD~(\ref{msd s}) where we observe the characteristic crossover dynamics from $\langle x^2(t)\rangle\sim t^{2p}$ to $\langle x^2(t)\rangle\sim t^{p}$. 
%{\color{blue}{We note that there are no logarithmic corrections to the MSD for some limiting cases of $\alpha$ since the diffusion coefficient in the general form of the MSD~(\ref{msd s}) has no divergence for any $\alpha<2$. The special case with $\alpha=2$ needs separate treatment and is not considered in this paper, as it was stated before.}}

\begin{table}[]
    \centering
    \begin{tabular}{|c|c|c|}
         \hline
         & \,MSD -- short time behavior\, & \, MSD -- long time behavior\, \\ 
         \hline
         & \,$\langle x^2(t) \rangle \sim t^{\mu_1}$, $\mu_1=4/(2-\alpha)$\, & \,$\langle x^2(t) \rangle \sim t^{\mu_2}$, $\mu_2=\mu_1/2$\, \\
         
         \hline
         \,$1<\alpha<2$\, & \,$\mu_1>4$ -- hyperdiffusion\, & \,$\mu_2>2$ -- hyperdiffusion\, \\
         \hline
         \,$\alpha=1$\, & \,$\mu_1=4$ -- hyperdiffusion\, & \,$\mu_2=2$ -- ballistic motion\, \\
         \hline
         \,$0<\alpha<1$\, & \,\,$2<\mu_1<4$ -- hyperdiffusion\,\, & \,\,$1<\mu_2<2$ -- superdiffusion\,\, \\
         \hline
         \,$\alpha=0$\, & \,$\mu_1=2$ -- ballistic motion\, & \,$\mu_2=1$ -- normal diffusion\, \\
         \hline
         \,\,$-2<\alpha<0$\,\, & \,\,$1<\mu_1<2$ -- superdiffusion\,\, & \,\,$1/2<\mu_2<1$ -- subdiffusion\,\, \\
         \hline
         \,$\alpha=-2$\, & \,$\mu_1=1$ -- normal diffusion\, & \,$\mu_2=1/2$ -- subdiffusion\, \\
         \hline
         \,$\alpha<-2$\, & \,$0<\mu_1<1$ -- subdiffusion\, & \,\,$0<\mu_2<1/2$ -- subdiffusion\,\, \\
         \hline
    \end{tabular}
    \caption{Characteristic crossover regimes in finite-velocity HDPs}\label{tab:my_label}
\end{table}

\section{Finite-velocity HDP with stochastic resetting}

\subsection{Probability density function and non-equilibrium stationary state}

We now turn to the analysis of the effect of stochastic resetting on the finite-velocity HDP. We consider a Poissonian stochastic resetting mechanism~\cite{evans1,evans2} with instantaneous resetting events, whereas we leave the case of non-instantaneous resetting for future research. Thus from the simple renewal equation approach we deduce~\cite{evans2014,mendez2019,bodrova1,bodrova2}
\begin{align}\label{renewal eq}
    P_{r}(x,t)=e^{-rt}P(x,t) + \int_{0}^{t}r\,e^{-rt'}P(x,t')\,dt',
\end{align}
where $P(x,t)$ is the PDF~(\ref{final pdf}) in absence of resetting, which means that each resetting event to the initial position $x_0$ renews the process at a rate $r$. In this equation, the first term on the right-hand side corresponds to the fraction that there is no resetting event up to time $t$, while the second term describes multiple resetting events up to time $t$. %{\color{blue}{Here we note that the resetting is instantaneous, i.e., with infinite velocity, which makes the model analytically treatable. The more realistic models with noninstantaneous resetting we leave for future research.}} 
The standard telegrapher's equation, which is a special case of our model for $p=1$ ($\alpha=0$), in presence of stochastic resetting was analysed in~\cite{masoliverreset,rt1}. Numerous examples of space-dependent diffusion in soft matter systems and recent experimental advances~\cite{exper} motivated new studies on inhomogeneous diffusion processes with resetting. Thus, the particular cases of heterogeneous diffusion processes with $\alpha = 1$ and stochastic Poissonian resetting were considered~\cite{ray1,ray2}. In this context, our study is a natural generalization of these very recent advances. Moreover, in view of the importance of resetting phenomena in the context of search problems, we can speculate that the heterogeneous telegrapher's equation with resetting represents a toy model of random search in a turbulent environment. We also note that run-and-tumble particle motion under stochastic resetting \cite{rt1,rt2,rt3,rt4} and more generalized models of L\'evy walks under resetting~\cite{xu,xu2} are of current interest.

By Laplace transform of equation~(\ref{renewal eq}), it follows that
\begin{align}\label{renewal eq laplace}
    \hat{P}_{r}(x,s)=\frac{s+r}{s}\hat{P}(x,s+r).
\end{align}
Using this relation, and in combination with equation~(\ref{cattaneo eq_general laplace}), we arrive at the relation
\begin{align}\label{cattaneo eq_general laplace_Pr} 
\tau\left[s^2\hat{P}_r(x,s)-s\delta(x-x_0)\right] &+ (2r\tau+1)\left[s\hat{P}_r(x,s)-\delta(x-x_0)\right]\nonumber\\ &= \mathcal{D}_{\alpha}\frac{\partial}{\partial x}\left\{|x|^{\frac{\alpha}{2}}\frac{\partial}{\partial x}\left[|x|^{\frac{\alpha}{2}}\hat{P}_r(x,s)\right]\right\} - r(r\tau+1)\left[\hat{P}_r(x,s)-\frac{1}{s}\delta(x-x_0)\right].
\end{align}
By inverse Laplace transform we derive the corresponding telegrapher's equation with position-dependent diffusion coefficient in the presence of stochastic resetting,
\begin{align}\label{cattaneo eq heterogeneous fpe} 
\tau\frac{\partial^{2}}{\partial t^{2}}P_r(x,t)+(2r\tau+1)\frac{\partial}{\partial t}P_r(x,t)=\mathcal{D}_{\alpha}\frac{\partial}{\partial x}\left\{|x|^{\frac{\alpha}{2}}\frac{\partial}{\partial x}\left[|x|^{\frac{\alpha}{2}}P_r(x,t)\right]\right\} - r(r\tau+1)\left[P_r(x,t)-\delta(x-x_0)\right].
\end{align}

\begin{figure}[t!]
\centering{(a)\includegraphics[width=8cm]{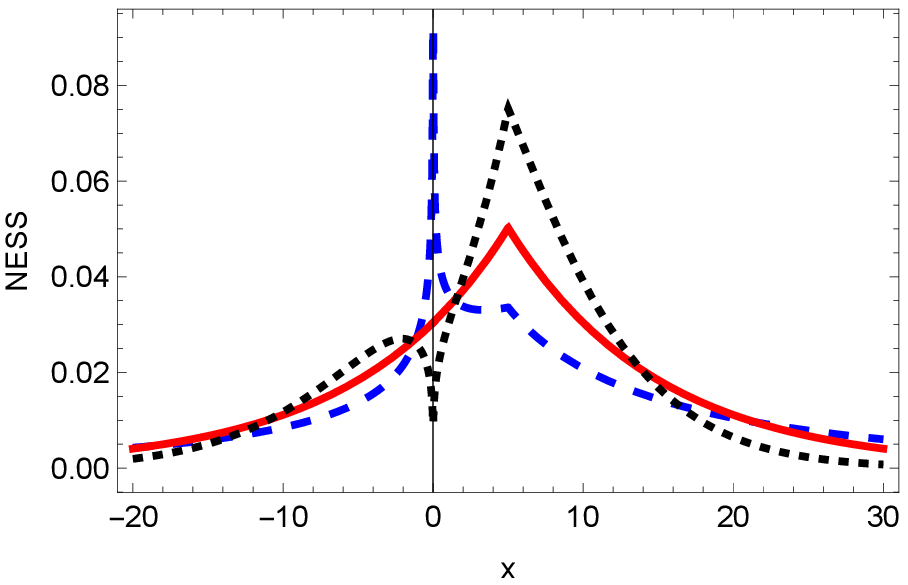} (b)\includegraphics[width=8cm]{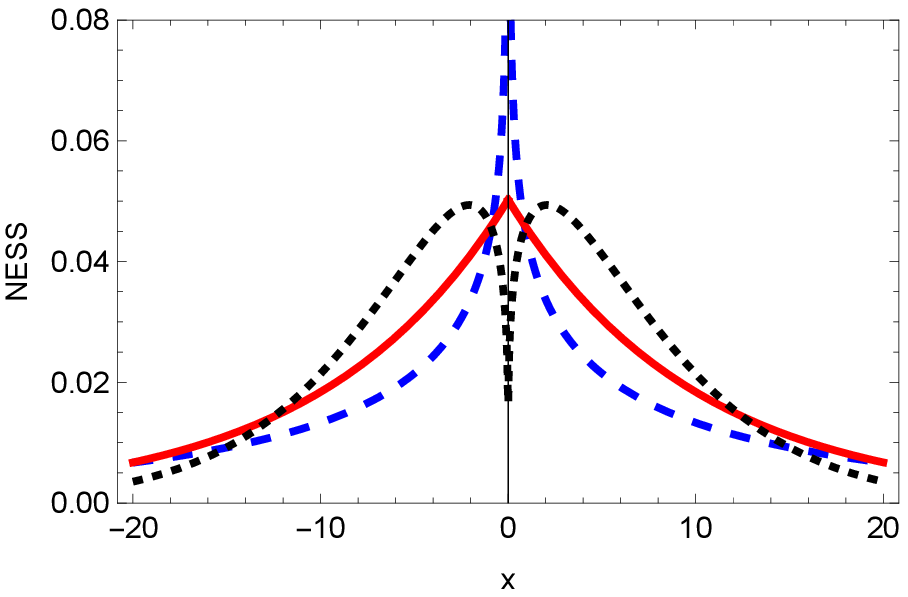} 
\caption{NESS (\ref{pdf reset ness}) for $\tau=1$, $v_\alpha=1$, $r=0.01$ and $\alpha=0.5$ (blue dashed line), $\alpha=0$ (red solid line), $\alpha=-0.5$ (black dotted line). (a) $x_0=5$, (b) $x_0=0$.}\label{fig_ness}}
\end{figure}

Next, we will show that in the long time limit the system reaches a non-equilibrium stationary state (NESS). From Eqs.~(\ref{pdf laplace final}) and (\ref{renewal eq laplace}), we obtain
\begin{align}\label{pdf laplace reset}
    \hat{P}_{r}(x,s)=&\frac{|x|^{1/p-1}}{2v_p}\frac{(s+r+\tau^{-1})^{1/2}(s+r)^{1/2}}{s}\nonumber\\&\times \exp\left(-\frac{p}{v_p}\left|\text{sgn}(x)|x|^{1/p}-\text{sgn}(x_0)|x_0|^{1/p}\right|(s+r+\tau^{-1})^{1/2}(s+r)^{1/2}\right).
\end{align}
From here, from the final value theorem $\lim_{t\rightarrow\infty}f(t)=\lim_{s\rightarrow0}s\hat{f}(s)$~\cite{schiff}, we find the NESS
\begin{align}\label{pdf reset ness}
    P_{r}^{st}(x)&=\lim_{t\rightarrow\infty}P_{r}(x,t)=\lim_{s\rightarrow0}s\hat{P}_{r}(x,s)\nonumber\\&=\frac{|x|^{1/p-1}}{2v_p}\sqrt{r(r+\tau^{-1})}\times \exp\left(-\sqrt{r(r+\tau^{-1})}\frac{p}{v_p}\left|\text{sgn}(x)|x|^{1/p}-\text{sgn}(x_0)|x_0|^{1/p}\right|\right).
\end{align}
For $\tau\rightarrow0$ (note that $v_p=\sqrt{\mathcal{D}_p/\tau}\rightarrow\infty$), we recover the result for HDPs with stochastic resetting, see equation~(26) in Ref.~\cite{hdpjpa2022},
\begin{align}\label{pdf reset ness tau=0}
    p_{r}^{st}(x)=\frac{|x|^{1/p-1}}{2\sqrt{\mathcal{D}_p/r}}\times \exp\left(-\frac{p}{\sqrt{\mathcal{D}_p/r}}\left|\text{sgn}(x)|x|^{1/p}-\text{sgn}(x_0)|x_0|^{1/p}\right|\right).
\end{align}
A graphical representation of the NESS is given in figure~\ref{fig_ness}. For $\alpha=0$ ($p=1$) this is in fact a Laplace distribution~\cite{masoliverreset,rt1}
\begin{align}\label{laplace distribution}
    P_{r}^{st}(x)=\frac{\sqrt{r(r+\tau^{-1})}}{2v}\times \exp\left(-\frac{\sqrt{r(r+\tau^{-1})}}{v}\left|x-x_0\right|\right).
\end{align}
In figure~\ref{fig_ness}(a) we observe that the PDF has a cusp at $x_0>0$ since the resetting mechanism introduces a source of probability at $x_0$. For $\alpha>0$ we observe another cusp at $x=0$ since for small $x$ the intensity of the multiplicative noise in the Langevin equation becomes very small such that the particle spends more time around the origin before it is reset to the initial position $x_0$. For $\alpha<0$ the PDF shows an anti-cusp at $x=0$ since for small $x$ the intensity of the multiplicative noise becomes
very large, and the particle does not spend much time near the origin. For $x_0=0$, see figure~\ref{fig_ness}(b), the PDFs show a cusp only at $x=0$.

\subsection{Transition to the non-equilibrium stationary state}

In order to find the relaxation dynamics to the NESS, we consider the renewal equation (\ref{renewal eq}). We see that in the long time limit the dominant term is the integral term, which will be estimated by the Laplace approximation for large $t$. For the integral, we obtain (see Appendix~\ref{app_a})
\begin{align}\label{renewal integral}
    \int_{0}^{t}r\,e^{-rt'}P(x,t')\,dt'\sim r\frac{|x|^{1/p-1}}{4v_p\sqrt{\tau\pi}}\sqrt{t}\int_{0}^{1}\left[1+\frac{\tau_{0}}{\sqrt{\tau_{0}^{2}-\frac{w^2}{v_p^2}}}\right]\frac{e^{-t\Phi(\tau_{0},w)}}{\left(\tau_{0}^{2}-\frac{w^2}{v_p^2}\right)^{1/4}}d\tau_{0},
\end{align}
where
\begin{align}
    \Phi(\tau_{0},w)=\left(r+\frac{1}{2\tau}\right)\tau_{0}-\frac{1}{2\tau}\sqrt{\tau_{0}^{2}-\frac{w^2}{v_p^2}},
\end{align}
and $$w=\frac{p\left|\text{sgn}(x)|x|^{1/p}-\text{sgn}(x_0)|x_0|^{1/p}\right|}{t}.$$Note that the integral in (\ref{renewal integral}) is always convergent in spite of the singularity arising from the denominator in the integrand. From the Laplace approximation \cite{arfken}
\begin{align}
    \mathcal{I}(t)\approx e^{-t\,f(z_0)}g(z_0)\sqrt{\frac{2\pi}{t|f''(z_0)|}},
\end{align}
of the integral
\begin{align}
    \mathcal{I}(t)=\int_0^1e^{-tf(z)}g(z)\,dz
\end{align}
for large $t$, which requires the evaluation of the minimum of the function $f(z)$, i.e., $f'(z_0)=0$, if $0<z_0<1$ (if the extremum point $z_0$ is outside the integration limits, $z_0>1$, then the approximation result is calculated at $z_0=1$), we find the integral (\ref{renewal integral}). The extremum point can be calculated from $\left.\frac{\partial}{\partial \tau_{0}}\Phi(\tau_0,w)\right|_{\tau_0=\tau_{0}^{\ast}}=0$, which gives
\begin{align}
    \tau_{0}^{\ast}=\frac{2r\tau+1}{\sqrt{(2r\tau+1)^2-1}}\frac{w}{v_p}.
\end{align}
From here, we find that the PDF behaves as
\begin{align}
    P_{r}(x,t)\sim e^{-t\,I(w)},
\end{align}
where the large deviation function (LDF) reads
\begin{align}\label{ldf}
    I(w)=\left\lbrace\begin{array}{ll}
        \sqrt{r(r+\tau^{-1})}\frac{w}{v_p}, & \quad \left|\text{sgn}(x)|x|^{1/p}-\text{sgn}(x_0)|x_0|^{1/p}\right|\le\frac{2\sqrt{r\tau(r\tau+1)}}{p(2r\tau+1)}v_{p}t, \\ \\
        \left(r+\frac{1}{2\tau}\right)-\frac{1}{2\tau}\sqrt{1-\frac{w^2}{v_p^2}}, & \quad \left|\text{sgn}(x)|x|^{1/p}-\text{sgn}(x_0)|x_0|^{1/p}\right|\ge\frac{2\sqrt{r\tau(r\tau+1)}}{p(2r\tau+1)}v_{p}t.
    \end{array}\right.
\end{align}
From here we conclude that the length scale grows like $\xi(t)\sim \left(v_pt\right)^{p}$. 

\begin{figure}[t!]
\centering{(a)\includegraphics[width=8cm]{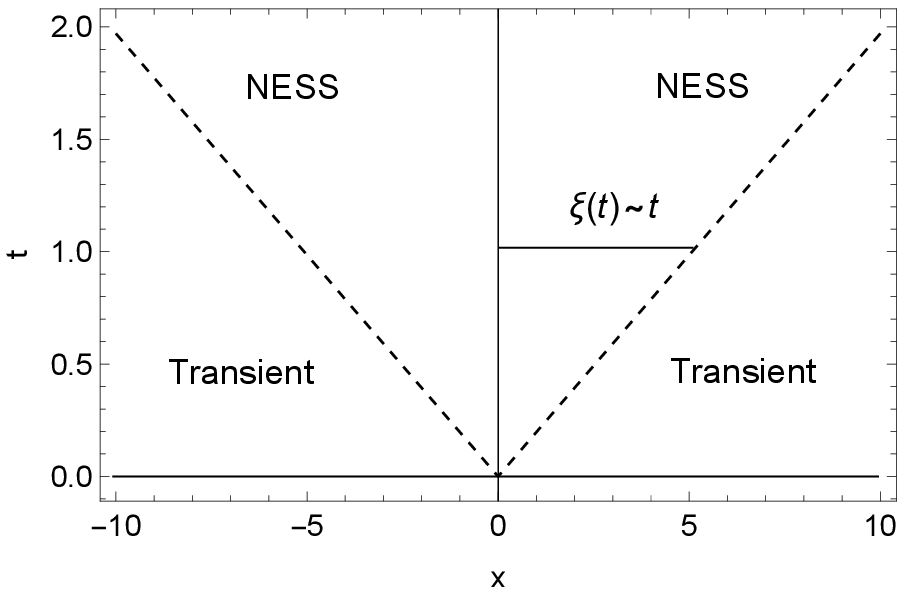} \\
\medskip 
(b)\includegraphics[width=8cm]{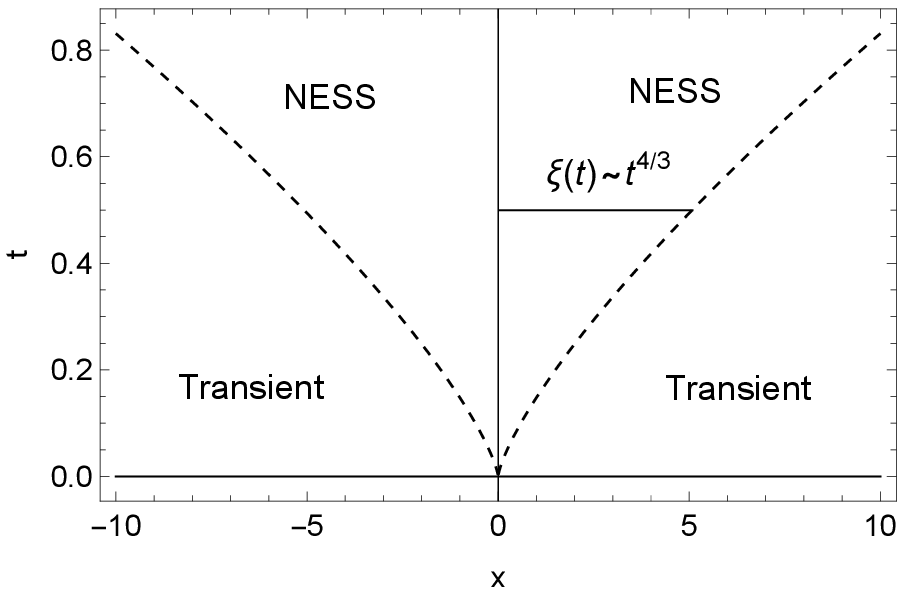} (c)\includegraphics[width=8cm]{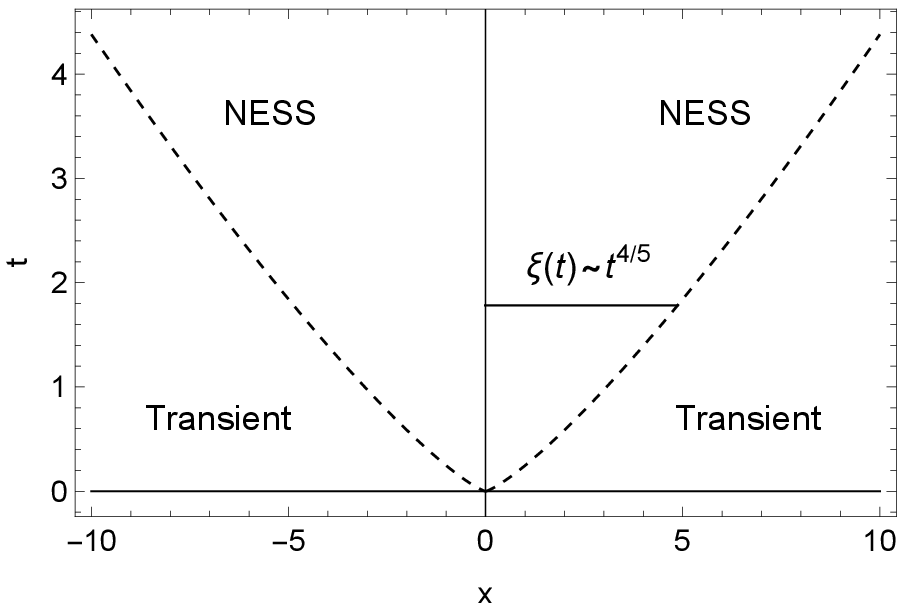}
\caption{Boundary between the region where the NESS is achieved and the transient region for $\tau=1$, $v_\alpha=1$, $r=0.01$ and (a) $\alpha=0$ -- standard telegrapher's process, (b) $\alpha=0.5$, (c) $\alpha=-0.5$.}\label{fig_ness_regions}}
\end{figure}

For $x_0=0$, we have ($w=p\frac{|x|^{1/p}}{t}$)
\begin{align}\label{ldf2}
    I(|x|/\xi(t))=\left\lbrace\begin{array}{ll}
        p\sqrt{r(r+\tau^{-1})}\frac{(|x|/t^p)^{1/p}}{v_p}, & \quad |x|\le\left(\frac{2\sqrt{r\tau(r\tau+1)}}{p(2r\tau+1)}\right)^pv_{p}^{p}t^{p}, \\ \\
        \left(r+\frac{1}{2\tau}\right)-\frac{1}{2\tau}\sqrt{1-p^2\frac{(|x|/t^p)^{2/p}}{v_p^2}}, & \quad |x|\ge\left(\frac{2\sqrt{r\tau(r\tau+1)}}{p(2r\tau+1)}\right)^pv_{p}^{p}t^{p},
    \end{array}\right.
\end{align}
and the length scale is $\xi(t)\sim \left(v_pt\right)^p$. The trajectories corresponding to the first line of equation~(\ref{ldf2}) are relaxed to the NESS (note that the LDF corresponds to the one of the PDF~(\ref{pdf reset ness}), as it should), while those satisfying the second line of equation~(\ref{ldf2}) are not relaxed and are still in transient regime. The boundary between the NESS region and the transient region moves with a non-constant velocity $v(t)\sim v_p^pt^{p-1}$ (see figure~\ref{fig_ness_regions}). For $\tau\rightarrow0$ the LDF reduces to the one for the HDP~\cite{hdpjpa2022}. For the standard telegrapher's equation ($p=1$, i.e., $\alpha=0$), the LDF becomes
\begin{align}\label{ldf2_standard}
    I(|x|/t)=\left\lbrace\begin{array}{ll}
        \sqrt{r(r+\tau^{-1})}\frac{|x|/t}{v}, & \quad |x|\le\frac{2\sqrt{r\tau(r\tau+1)}}{2r\tau+1}vt, \\ \\
        \left(r+\frac{1}{2\tau}\right)-\frac{1}{2\tau}\sqrt{1-\frac{x^2/t^2}{v^2}}, & \quad |x|\ge\frac{2\sqrt{r\tau(r\tau+1)}}{2r\tau+1}vt.
    \end{array}\right.
\end{align}
For $\tau\rightarrow0$ (HDP), we arrive at the known LDF \cite{ldf_paper}
\begin{align}
    I(|x|/t)=\left\lbrace\begin{array}{ll}
        \sqrt{r/D}\,\frac{|x|}{t}, & |x|\leq\sqrt{4Dr}\,t, \\ \\
        r+\frac{1}{4D}\left(\frac{|x|}{t}\right)^{2}, & |x|\ge\sqrt{4Dr}\,t.
    \end{array}\right.
\end{align}
In figure~\ref{fig_ness_regions}, we show the boundaries between the regions in which the particles have already relaxed to the NESS and the region in which the particles are in a transient regime. It is evident that the length scale depends on the parameter $\alpha$. 

\subsection{Mean squared displacement}

The MSD can be calculated from equation~(\ref{renewal eq laplace}), yielding
\begin{align}\label{renewal eq laplace msd}
    \langle x^2(s)\rangle_{r} =\frac{s+r}{s}\langle \hat{x}^2(x,s+r)\rangle_{\tau},
\end{align}
where $\langle \hat{x}^2(x,s)\rangle_{\tau}$ is given by equation~(\ref{msd laplace te hdp}). From here we conclude that in the short time limit ($s\rightarrow\infty$) the MSD in presence of resetting behaves analogously to the MSD in absence of resetting $\langle x^2(t)\rangle_{r} =\langle \hat{x}^2(x,t)\rangle_{\tau}$, while in the long time limit ($s\rightarrow0$) it saturates to $\langle x^2(t)\rangle_{r} = r\langle \hat{x}^2(x,r)\rangle_{\tau}$.

\begin{figure}[t!]
\centering{\includegraphics[width=8cm]{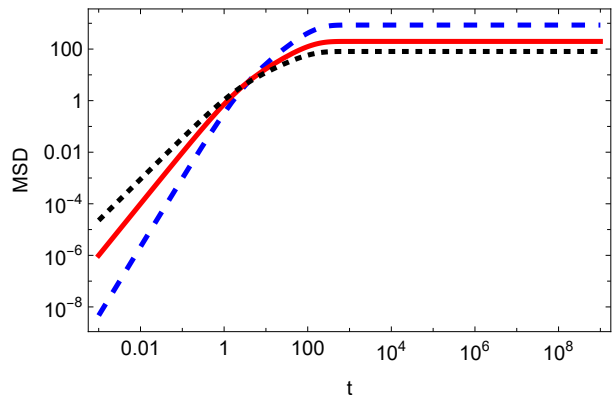}
\caption{MSD~(\ref{msd s reset x0=0}) for $\tau=1$, $\mathcal{D}_\alpha=1$, $r=0.01$, and $\alpha=0.5$ (blue dashed line), $\alpha=0$ (red solid line), $\alpha=-0.5$ (black dotted line).}\label{fig_msd_reset}}
\end{figure}

For $x_0=0$, the MSD becomes
\begin{align}\label{msd s reset x0=0}
\left\langle x^{2}(t)\right\rangle_{r}
&=\frac{\Gamma\left(1+2p\right)\left(\mathcal{D}_{\alpha}/\tau\right)^{p}}{p^{2p}}\mathcal{L}^{-1}\left[\frac{s^{-1}(s+r)^{-p}}{(s+r+\tau^{-1})^{p}}\right]\nonumber\\&=\frac{\Gamma\left(1+2p\right)\left(\mathcal{D}_{\alpha}\tau\right)^{p}}{p^{2p}\tau}\int_{0}^{t}e^{-rt'}\left(\frac{t'}{\tau}\right)^{2p-1}E_{1,2p}^{p}\left(-\frac{t'}{\tau}\right)\,dt'.
\end{align}
In the short time limit it behaves as the MSD in absence of resetting, equation~(\ref{msd s}),
\begin{align}\label{msd s reset x0=0 short time}
\left\langle x^{2}(t)\right\rangle_{r}
&\sim\frac{\Gamma\left(1+2p\right)\left(\mathcal{D}_{\alpha}\tau\right)^{p}}{p^{2p}\tau}\int_{0}^{t}\left(\frac{t'}{\tau}\right)^{2p-1}E_{1,2p}^{p}\left(-\frac{t'}{\tau}\right)\,dt'\nonumber\\&=\frac{\Gamma\left(1+2p\right)\left(\mathcal{D}_{\alpha}\tau\right)^{p}}{p^{2p}}\left(\frac{t}{\tau}\right)^{2p}E_{1,2p+1}^{p}\left(-\frac{t}{\tau}\right),
\end{align}
while in the long time limit the MSD saturates to $\langle x^{2}(t)\rangle_{r}\sim \frac{1}{r^{p}(r+\tau^{-1})^{p}}$, due to the resetting mechanism. A graphical representation of the MSD~(\ref{msd s reset x0=0}) is shown in figure~\ref{fig_msd_reset}, where the its saturation due to the stochastic resetting is clearly observed.

\section{Summary}

We reported exact results for the heterogeneous telegrapher's equation. A rich range of different diffusion regimes were observed, such as a crossover from hyperdiffusion to either superdiffusion, ballistic motion, or hyperdiffusion with different exponent, from ballistic motion to normal diffusion, from superdiffusion to subdiffusion, from normal diffusion to subdiffusion, or from subdiffusion with larger exponent to subdiffusion with lower exponent. Therefore, the considered model is suitable to describe anomalous diffusion in complex systems exhibiting characteristic crossover dynamics, including finite-velocity diffusion in random media. We also analysed the finite-velocity HDP with stochastic resetting and we showed that the system reaches a NESS. The transition to the NESS was analysed in terms of the large deviation function. We also found the boundaries between the region in which the system relaxed to the NESS and the transient region as a function of $\alpha$. Exact results for the MSD under resetting were obtained, as well. The anomalous diffusive regime saturates in the long time limit due to the resetting mechanism. 

Future research could be related to the investigation of ergodic properties of finite-velocity HDPs in absence and presence of resetting \cite{chchme,ch1,ch3,pregbm,jpagbm,wei}, including also corresponding higher-dimensional formulations~\cite{ch2,me2020}. Infinite- and finite-velocity HDPs in presence of time-dependent resetting~\cite{pal_t}, non-instantaneous~\cite{bodrova2,radice1} and space-time coupled returns~\cite{pal_njp}, HDPs in presence of resetting in an interval~\cite{pal_int,roldan} and bounded in complex potential~\cite{csf_reset}, as well as discrete space-time resetting models~\cite{discrete models} for HDPs, are other topics worth investigating.

\section*{Acknowledgements}
{TS was supported by the Alexander von Humboldt Foundation. AC acknowledges support of the Polish National Agency for Academic Exchange (NAWA). TS and LK also acknowledge support from the bilateral Macedonian-Chinese research project 20-6333, funded under the intergovernmental Macedonian-Chinese agreement. The authors acknowledge financial support by the German Science Foundation (DFG, Grant No. ME 1535/12-1).}

\appendix

\section{Different forms of heterogeneous telegrapher's equation}\label{app_different forms}

Here we note that one can derive different form of the heterogeneous telegrapher's equations for a voltage and current in a lossy transmission inhomogeneous line~\cite{book}, which have the form
\begin{align}
    &\tau\frac{\partial^2}{\partial t^2}I(x,t)+\frac{\partial}{\partial t}I(x,t)=\frac{\partial}{\partial x}\left[D(x)\frac{\partial}{\partial x}I(x,t)\right],\label{eq_I}\\
    &\tau\frac{\partial^2}{\partial t^2}V(x,t)+\frac{\partial}{\partial t}V(x,t)=D(x)\frac{\partial^2}{\partial x^2}V(x,t),\label{eq_V}
\end{align}
where $\tau=L/R=\text{const}$, $R=\text{const}$ is the resistance, $L=\text{const}$ is the inductance, $D(x)=\left[RC(x)\right]^{-1}$, and $C(x)$ is the capacitance.

Telegrapher's equation of form (\ref{eq_I}) can also be derived from the continuity equation
\begin{align}\label{con eq}
    \frac{\partial}{\partial t}n(x,t)+\frac{\partial}{\partial x}J(x,t)=0,
\end{align}
where $n(x,t)$ is the concentration and the flow of particles $J(x,t)$ obeys the generalized Fick's law with memory,
\begin{align}\label{gFL}
    J(x,t)=\frac{1}{\tau}\int_{0}^{t}e^{(t-t')/\tau}D(x)\frac{\partial}{\partial x}n(x,t')\,dt',
\end{align}
where $\tau$ is a time parameter. From equations~(\ref{con eq}) and (\ref{gFL}), one arrives at the heterogeneous telegrapher's equation
\begin{align}
    \tau\frac{\partial^2}{\partial t^2}n(x,t)+\frac{\partial}{\partial t}n(x,t)=\frac{\partial}{\partial x}\left[D(x)\frac{\partial}{\partial x}n(x,t)\right].
\end{align}

Another form of the heterogeneous telegrapher's equation can be derived from the persistent random walk in inhomogeneous medium. It takes the form \cite{weiss2002,masoliver1994,masoliverweiss}
\begin{align}
    \frac{\partial^2}{\partial t^2}P(x,t)+\frac{1}{\tau}\frac{\partial}{\partial t}P(x,t)=v(x)\frac{\partial}{\partial x}\left[v(x)\frac{\partial}{\partial x}P(x,t)\right],
\end{align}
where $v(x)$ is the position-dependent velocity.

\section{Normalization of the PDF}\label{app norm}

Here we provide detailed proof of the normalization condition $\int_{-\infty}^{\infty}P(x,t)\,dx=1$. From Eq.~(\ref{P through L}), we have
\begin{align}\label{P through L norm}
    \int_{-\infty}^{\infty}P(x,t)\,dx=\mathcal{J}+\tau\,\frac{\partial}{\partial t}\mathcal{J},
\end{align}
with
\begin{align}
    \mathcal{J}=\frac{e^{-\frac{t}{2\tau}}}{2v_{p}\tau}\int_{-\infty}^{\infty}|x|^{1/p-1}&\theta\left(v_p t-p\left|\text{sgn}(x)|x|^{1/p}-\text{sgn}(x_0)|x_0|^{1/p}\right|\right)
     \nonumber\\& \times I_{0}\left(\frac{\sqrt{v_p^2t^2-p^2\left[\text{sgn}(x)|x|^{1/p}-\text{sgn}(x_0)|x_0|^{1/p}\right]^2}}{2v_p \tau}\right)dx=\mathcal{J}_1+\mathcal{J}_2,
\end{align}
where
\begin{align}
    \mathcal{J}_1=\frac{e^{-\frac{t}{2\tau}}}{2v_{p}\tau}\int_{0}^{\infty}|x|^{1/p-1}&\theta\left(v_p t-p\left|\text{sgn}(x)|x|^{1/p}-\text{sgn}(x_0)|x_0|^{1/p}\right|\right)
     \nonumber\\& \times I_{0}\left(\frac{\sqrt{v_p^2t^2-p^2\left[\text{sgn}(x)|x|^{1/p}-\text{sgn}(x_0)|x_0|^{1/p}\right]^2}}{2v_p \tau}\right)dx
\end{align}
and
\begin{align}
    \mathcal{J}_2=\frac{e^{-\frac{t}{2\tau}}}{2v_{p}\tau}\int_{-\infty}^{0}|x|^{1/p-1}&\theta\left(v_p t-p\left|\text{sgn}(x)|x|^{1/p}-\text{sgn}(x_0)|x_0|^{1/p}\right|\right)
     \nonumber\\& \times I_{0}\left(\frac{\sqrt{v_p^2t^2-p^2\left[\text{sgn}(x)|x|^{1/p}-\text{sgn}(x_0)|x_0|^{1/p}\right]^2}}{2v_p \tau}\right)dx.
\end{align}
For the first integral, by introducing $p\,x^{1/p} = y$, i.e., $x^{1/p-1}dx = dy$ and then $z=y-p\,\text{sgn}(x_0)|x_0|^{1/p}$, i.e., $dy=dz$, we find
\begin{align}
    \mathcal{J}_1&=\frac{e^{-\frac{t}{2\tau}}}{2v_{p}\tau}\int_{0}^{\infty}\theta\left(v_p t-\left|y-p\,\text{sgn}(x_0)|x_0|^{1/p}\right|\right) I_{0}\left(\frac{\sqrt{v_p^2t^2-\left[y-p\,\text{sgn}(x_0)|x_0|^{1/p}\right]^2}}{2v_p \tau}\right)dy\nonumber\\&
    =\frac{e^{-\frac{t}{2\tau}}}{2v_{p}\tau}\int_{0}^{\infty}\theta\left(v_p t-\left|z\right|\right) I_{0}\left(\frac{\sqrt{v_p^2t^2-z^2}}{2v_p \tau}\right)dz.
\end{align}
Then we introduce new variable $\frac{\sqrt{v_p^2t^2-z^2}}{2v_p\,t}=r$ to obtain
\begin{align}
    \mathcal{J}_1=\frac{e^{-\frac{t}{2\tau}}}{2v_p\,\tau}\int_{0}^{\frac{t}{2\tau}}I_{0}(r)\frac{r\,dr}{\sqrt{\left(\frac{t}{2\tau}\right)^2-r^2}}=e^{-\frac{t}{2\tau}}\sinh\left(\frac{t}{2\tau}\right)
\end{align}
For the second integral, we first introduce $z = -x$ and then $p\,z^{1/p} = y$ and $k=y+p\,\text{sgn}(x_0)|x_0|^{1/p}$, to find
\begin{align}
    \mathcal{J}_2&=\frac{e^{-\frac{t}{2\tau}}}{2v_{p}\tau}\int_{0}^{\infty}\theta\left(v_p t-\left|y+p\,\text{sgn}(x_0)|x_0|^{1/p}\right|\right) I_{0}\left(\frac{\sqrt{v_p^2t^2-\left[y+p\,\text{sgn}(x_0)|x_0|^{1/p}\right]^2}}{2v_p \tau}\right)dy\nonumber\\&
    =\frac{e^{-\frac{t}{2\tau}}}{2v_{p}\tau}\int_{0}^{\infty}\theta\left(v_p t-\left|k\right|\right) I_{0}\left(\frac{\sqrt{v_p^2t^2-k^2}}{2v_p \tau}\right)dk=e^{-\frac{t}{2\tau}}\sinh\left(\frac{t}{2\tau}\right).
\end{align}
Therefore, $\mathcal{J}=2e^{-\frac{t}{2\tau}}\sinh\left(\frac{t}{2\tau}\right)=1-e^{-\frac{t}{\tau}}$. Thus, we finaly obtain
\begin{align}\label{P norm final}
    \int_{-\infty}^{\infty}P(x,t)\,dx=\left(1-e^{-\frac{t}{\tau}}\right)+\tau\,\frac{\partial}{\partial t}\left(1-e^{-\frac{t}{\tau}}\right)=1-e^{-\frac{t}{\tau}}+e^{-\frac{t}{\tau}}=1,
\end{align}
which completes the proof.

\section{Calculation of the integral in the renewal equation}\label{app_a}

Let us analyse the renewal equation
\begin{align}\label{renewal eq app}
    P_{r}(x,t)=e^{-rt}P(x,t) + \int_{0}^{t}r\,e^{-rt'}P(x,t')\,dt'.
\end{align}
For large time $t$ the integral term is dominant, and thus
\begin{align}\label{renewal eq app large t}
    P_{r}(x,t)\sim \int_{0}^{t}r\,e^{-rt'}P(x,t')\,dt'.
\end{align}
We introduce 
\begin{align}
    \xi'=\frac{\sqrt{t'^2-\frac{X^2}{v_\alpha^2}}}{2\tau},
\end{align}
where $$X=p\left|\text{sgn}(x)|x|^{1/p}-\text{sgn}(x_0)|x_0|^{1/p}\right|.$$ We also use $t'=t\tau_{0}$ ($dt'=t\,d\tau_{0}$), from where it follows
\begin{align}
    \xi'=\frac{t}{2\tau}\sqrt{\tau_{0}^2-\frac{w^2}{v_\alpha^2}},
\end{align}
where $w^2=X^2/t^2$. Since when $t$ is large then $\xi'$ is also large, and the Bessel function behaves as $I_{\nu}(\xi')\sim \frac{e^{\xi'}}{\sqrt{2\pi\xi'}}$. Thus, for the integral, we obtain
\begin{align}\label{renewal integral app}
    \int_{0}^{t}r\,e^{-rt'}P(x,t')\,dt'&\sim \int_{0}^{t}r\,e^{-rt'}\frac{|x|^{1/p-1}}{4v_\alpha\tau}e^{-\frac{t'}{2\tau}}\left[I_{0}(\xi')+\frac{t'}{2\tau}\frac{I_1(\xi')}{\xi'}\right]dt'\nonumber\\&\sim r\frac{|x|^{1/p-1}}{4v_\alpha\tau}\int_{0}^{t}e^{-\left(r+\frac{1}{2\tau}\right)t'}\left[1+\frac{t'}{2\tau}\frac{1}{\xi'}\right]\frac{e^{\xi'}}{\sqrt{2\pi\xi'}}dt'\nonumber\\&=r\frac{|x|^{1/p-1}}{4v_\alpha\tau}\int_{0}^{t}\left[1+\frac{t'}{2\tau}\frac{1}{\xi'}\right]\frac{1}{\sqrt{2\pi\xi'}}e^{-\left(r+\frac{1}{2\tau}\right)t'+\xi'}dt'\nonumber\\&=r\frac{|x|^{1/p-1}}{4v_\alpha\sqrt{\pi\tau}}\int_{0}^{t}\left[1+\frac{t'}{\sqrt{t'^2-\frac{X^2}{v_\alpha^2}}}\right]\frac{e^{-\left(r+\frac{1}{2\tau}\right)t'+\frac{1}{2\tau}\sqrt{t'^2-\frac{X^2}{v_\alpha^2}}}}{\left(t'^2-\frac{X^2}{v_\alpha^2}\right)^{1/4}}dt'\nonumber\\
    &=r\frac{|x|^{1/p-1}}{4v_\alpha\sqrt{\tau\pi}}\int_{0}^{1}\left[1+\frac{t\tau_{0}}{\sqrt{t^2\tau_{0}^{2}-\frac{X^2}{v_\alpha^2}}}\right]\frac{e^{-\left(r+\frac{1}{2\tau}\right)t\tau_{0}+\frac{t}{2\tau}\sqrt{\tau_{0}^{2}-\frac{w^2}{v_\alpha^2}}}}{\left(t^2\tau_{0}^{2}-\frac{X^2}{v_\alpha^2}\right)^{1/4}}td\tau_{0}\nonumber\\
    &=r\frac{|x|^{1/p-1}}{4v_\alpha\sqrt{\tau\pi}}\sqrt{t}\int_{0}^{1}\left[1+\frac{\tau_{0}}{\sqrt{\tau_{0}^{2}-\frac{w^2}{v_\alpha^2}}}\right]\frac{e^{-t\Phi(\tau_{0},w)}}{\left(\tau_{0}^{2}-\frac{w^2}{v_\alpha^2}\right)^{1/4}}d\tau_{0},
\end{align}
where
\begin{align}
    \Phi(\tau_{0},w)=\left(r+\frac{1}{2\tau}\right)\tau_{0}-\frac{1}{2\tau}\sqrt{\tau_{0}^{2}-\frac{w^2}{v_\alpha^2}}.
\end{align}

\end{document}